\documentclass[12pt, A4]{article}

\usepackage{amsmath}

\usepackage{amsfonts}
\usepackage{amssymb}
\usepackage[colorlinks,citecolor=blue,urlcolor=blue,bookmarks=false,hypertexnames=true]{hyperref}
\usepackage{cite}

\usepackage{graphicx, physics, subcaption}

\usepackage{color}

\usepackage{amsmath, amssymb, graphics, epsfig, graphicx}
\usepackage{epsf}
\usepackage{epstopdf}
\usepackage {amssymb}
\newcommand{\nc}{\newcommand}
\nc{\ba}{\begin{eqnarray}}
\nc{\ea}{\end{eqnarray}}

\def\bfk{{\bf k}}

\def\bfx{{\bf x}}

\nc{\cN}{ {\cal{N}} }

\begin{document}

%%%%%%%%%%%%%%%%%%%%%%%%%%%%%%%%%%%%%%%%%%%%%%%%%%%%%
%\begin{flushright} {\footnotesize IC/2007/001}  \end{flushright}
\vspace{5mm}
\vspace{0.5cm}
\begin{center}

\def\thefootnote{\fnsymbol{footnote}}

{\bf\large Stochastic Effects in Anisotropic Inflation }
\\[0.5cm]

{ Alireza Talebian$^1$ \footnote{talebian@ipm.ir}, 
Amin Nassiri-Rad$^1$ \footnote{amin.nassiriraad@ipm.ir}, 
Hassan Firouzjahi$^{1,2}$ \footnote{firouz@ipm.ir},
}
\\[0.5cm]
 
 {\small \textit{$^1$School of Astronomy, Institute for Research in Fundamental Sciences (IPM) \\ P.~O.~Box 19395-5531, Tehran, Iran
}}\\

{\small \textit{$^2$ Department of Physics, Faculty of Basic Sciences, University of Mazandaran, \\ 
P. O. Box 47416-95447, Babolsar, Iran}}\\

\end{center}

\vspace{.8cm}

\hrule \vspace{0.3cm}
%{\small  \noindent \textbf{Abstract} \\[0.3cm]
%\noindent

%%%%%%%%%%%%%%%%%%%%%%%%%%%%%%%%%%%%%%%%%%%%%%%%%%%%%
\begin{abstract}
We revisit the stochastic effects in the model of anisotropic inflation containing a $U(1)$ gauge field. 
We obtain the Langevin  equations for the inflaton and gauge fields  perturbations and solve them analytically. 
 We show that if the initial value of the electric field is larger than its classical attractor value, then the random stochastic forces associated with the gauge field  is balanced by the frictional damping (classical) force and the electric field falls into a equilibrium (stationary) regime. As a result,  the classical attractor value of the electric field is replaced by its stationary value. We show that the probability of generating quadrupolar statistical anisotropy consistent with  CMB constraints can be significant.

\end{abstract}

\vspace{0.5cm} \hrule
\def\thefootnote{\arabic{footnote}}
\setcounter{footnote}{0}
\newpage

%\tableofcontents
%%%%%%%%%%%%%%%%%%%%%%%%%%%%%%%%%%%%%%%%%%%%%%%%%%%%%
\section{Introduction}
\label{sec:intro}

Thanks to numerous cosmological observations, inflationary cosmology has emerged as the leading paradigm for the early universe cosmology. In its simplest realization, inflation is driven by a scalar field, the inflaton field, which slowly rolls overs its flat potential.  The quantum fluctuations of the inflaton field, stretched over the horizon of a near de-Sitter background, is the  origin of the Cosmic Microwave Background (CMB) temperature anisotropies and the seeds of the large scale structure.   The statistical features of these fluctuations are well consistent with the basic predictions of simple inflationary scenarios where the primordial perturbations are observed  to be  nearly scale invariant, adiabatic and almost Gaussian \cite{Akrami:2018odb,Ade:2015lrj}.

Although a statistically isotropic universe is well supported by the CMB observations, but the possibility of having ``anisotropic hair'' in the early universe has attracted considerable interests in recent years. The simplest mechanism to generate statistical anisotropies on large scales is based on the so called model of $f^2 F^2$ in which a $U(1)$ gauge field is non-minimally coupled to the inflaton field $\phi$ via the conformal coupling $f(\phi)$.  With an appropriate choice of the conformal coupling,  the energy is dragged continuously from the inflaton background to the gauge field sector such that  the background electric field survives the exponential expansion   while a nearly scale invariant gauge field perturbations can be generated on superhorizon scales \cite{Watanabe:2009ct}. The basic prediction of models of anisotropic inflation is that a quadrupolar type statistical anisotropy is generated on CMB maps. However, there are strong observational bounds on the amplitude of quadrupolar statistical anisotropy, $|g_*| \lesssim 0.01$ \cite{Ade:2015hxq, Kim:2013gka}, which poses strong constraint on the parameters of the models of anisotropic inflation. For various works related to anisotropic inflation see \cite{Watanabe:2009ct, Emami:2015qjl,  Watanabe:2010fh, Bartolo:2012sd, Emami:2013bk, Abolhasani:2013zya, Naruko:2014bxa, anisotropic-inflation}.

In order to meet the observational bounds on $g_*$, the contributions of the background electric field 
 to the total energy  must be kept under control. The energy density of the gauge field settles to an attractor value via the  back-reaction mechanism  \cite{Watanabe:2009ct}. Independent of the initial value of the background electric field energy density, the back-reactions between the gauge field and the inflaton field drive the system to its attractor regime in which the electric field energy density  is a small 
 but a nearly constant fraction of the total energy density.  However, this is the classical picture in which the  stochastic effects of the short wavelength perturbations are neglected. One might worry that the effects of short-wavelength perturbations  can significantly influence the background dynamics, preventing the system from reaching to its attractor solution \cite{ Fujita:2017lfu}. One of the well known approach to study quantum effects of these modes during inflation is the stochastic inflation formalism \cite{Starobinsky:1986fx, Linde:1986fd, Sasaki:1987gy, Nambu:1988je, Nakao:1988yi, Mollerach:1990zf, Starobinsky:1994bd, Finelli:2008zg, Finelli:2010sh, Enqvist:2011pt, Martin:2011ib, Kawasaki:2012bk, Fujita:2013cna, Fujita:2014tja, Vennin:2015hra, Tokuda:2017fdh,Vilenkin:1983xp, Nambu:1987ef,  Kandrup:1988sc,  Nambu:1989uf, Linde:1993xx, Kunze:2006tu, Prokopec:2007ak, Prokopec:2008gw, Tsamis:2005hd, Enqvist:2008kt, Garbrecht:2013coa, Garbrecht:2014dca, Burgess:2014eoa, Burgess:2015ajz, Boyanovsky:2015tba,  Boyanovsky:2015jen, Fujita:2017lfu,Vennin:2016wnk, Assadullahi:2016gkk, Grain:2017dqa, Noorbala:2018zlv}. In this approach, the quantum fluctuations are decomposed into the infrared (IR) and ultraviolet (UV) modes. As the  UV modes are stretched and leaving the Hubble horizon during inflation, they act as quantum kicks on super-horizon scale perturbations (classical solutions) with the amplitude $H/2\pi$ in which $H$ is Hubble expansion rate during inflation. These quantum noises can be translated into a stochastic force imposed on the classical slow-roll dynamics of the coarse grained IR modes. Consequently, the  deterministic interpretation of the dynamics of the field on large scale is replaced by  a probabilistic one.

Fujita and Obata \cite{Fujita:2017lfu} have studied the stochastic effects of the gauge fields perturbations 
 in the model of $f^2 F^2$ anisotropic inflation. They derived the Langevin  and the Fokker-Planck equations for the  gauge fields perturbations  and showed that the gauge field fluctuations pile up and quickly
overwhelm the classical attractor solution. By assuming that the  gauge field is in the stochastic equilibrium when the CMB modes exit the horizon, they reported that the probability of the CMB statistical anisotropy  to be consistent with the current observational upper bound is  around 0.001\%,  excluding $f^2F^2$ anisotropic inflation scenario with the probability about 99.999\%. It is argued  that this conclusion 
 is independent of the model parameters or the initial conditions.

In this paper we revisit the stochastic effects of the gauge field and the inflaton field  in anisotropic inflation model and derive the associated Langevin equations. Some of our results disagree with those of 
\cite{Fujita:2017lfu}. In particular, we show that  the fate of electric field is sensitive to its initial value $E_{\rm ini}$ relative to the classical attractor value $E_{\rm att}$. If $E_{\rm ini} \leq E_{\rm att}$,  the stochastic effects of the electric field become relevant only  after an exponentially large number of e-folds has passed. However,  for $E_{\rm ini} > E_{\rm att}$, the accumulative effects  of the stochastic noises are not strong enough to  spoil the nearly-isotropic approximation of $f^2F^2$ setup and the classical attractor mechanism of  \cite{Watanabe:2009ct} is reached. In this case the system falls into a equilibrium state where the stochastic and classical forces balance each other.

The paper is organized as follows. In Sec.~\ref{Aniso-Inf}, the  model of $f^2F^2$ anisotropic inflation  and its predictions for statistical anisotropy are reviewed. The Langevin equation of the electric field is derived in Sec.~\ref{stoc-rev} and the effects of quantum kicks on the coarse-grained electric field are studied in three subclasses. In Sec.~\ref{Scal_Sto}, the stochastic dynamics of the inflaton field are studied followed by the summaries and discussions  in Sec.~\ref{Summary}. The derivations of the correlation functions of the 
electric and magnetic noises  and a brief review of the probability distribution function are
presented in appendices \ref{Noise} and \ref{PDF}, respectively.
%%%%%%%%%%%%%%%%%%%%%%%%%%%%%%%%%%%%%%%%%%%%%%%%%%%%%
\section{Anisotropic Inflation}
\label{Aniso-Inf} 

Here we review the setup of anisotropic inflation and collect the necessary results for the stochastic analysis in next Sections.  The model contains the inflaton field $\phi$ which is a real scalar field  coupled to an Abelian gauge field $A_{\mu}$ through the gauge kinetic coupling  $f(\phi)$ via the following action
\begin{eqnarray}
		{S} = \int \mathrm{d}^4 x \, \sqrt{-g} \, \bigg[ \dfrac{M_{P}^2}{2} R - \dfrac{1}{2} g^{\mu \nu} \partial_{\mu} \phi  \partial_{\nu} \phi - V(\phi) - \dfrac{f^2(\phi)}{4} F^{\mu \nu} F_{\mu \nu} \bigg] \, ,
	\label{action}
\end{eqnarray}
in which $M_P$ is the reduced Planck mass,  $F_{\mu \nu}$ is the field strength associated with the $U(1)$ gauge field and  $R$ is the Ricci scalar. The background metric is in the form of  the Bianchi type I universe,
\begin{eqnarray}
\mathrm{d}s^2 = -\mathrm{d}t^2 + e^{2\alpha(t)} \, \hat{\gamma}_{ij}(t) \, \mathrm{d}x^i \mathrm{d}x^j.
\label{metric}
\end{eqnarray}
Here $\alpha$ is the number of $e$-folds so $\dot{\alpha} \equiv H$ represents the Hubble expansion rate, where a dot denotes the derivative with  respective to $t$.  Without loss of generality, we can align the $x$-axis along the direction of the background vector field, $\boldsymbol{A}=\big(A,0,0\big)$, and consider the  axisymmetric line element matrix  $\hat{\gamma}_{ij}$ as
\ba
\hat{\gamma}_{ij}(t)= \mathrm{diag} \Big(e^{-4\beta(t)}, e^{2\beta(t)},  e^{2\beta(t)}  \Big) \, ,
\ea
where $\dot{\beta}$ measures the anisotropy expansion rate. Observationally, the deviation from isotropy 
is very small, therefore we demand that the quantity $\Sigma \equiv \dfrac{\dot{\beta}}{H}$ measuring the fractional anisotropic expansion  to be very small, $\Sigma \ll 1$. 

Going to  Coulomb-radiation gauge $A_0 = 0 = \partial_i A_i$,  the equations of motion for the  inflaton and gauge fields respectively are 
\begin{eqnarray}
\ddot{\phi} - e^{-2\alpha}\nabla^2\phi + 3 H \dot{\phi} + V_{,\phi}(\phi) - \dfrac{f_{,\phi}(\phi)}{f(\phi)} (E^2+B^2) = 0 \,,\label{EoM-Phi} \\
\boldsymbol{\ddot{A}} - e^{-2\alpha}\nabla^2 \boldsymbol{A} + \bigg( H+2\dfrac{\dot{f}}{f} \bigg) \boldsymbol{\dot{A}} =0 \,. \label{EoM-A}
\end{eqnarray}
Hereafter, the bold symbols denote vector quantities. Associated with the gauge field $A_{i}$, we can define the  electric field $E_i \equiv -e^{-\alpha} f \partial_t A_i$ and the magnetic field $B_i \equiv e^{-2\alpha} \epsilon_{ijk} f \partial_j A_k$\,, where their magnitudes appear in Eq. \eqref{EoM-Phi}. Since the background is homogeneous, $B_i=0$ so  the energy density of the gauge field is given by the electric field,  
\begin{eqnarray}
\rho_E %&=& \dfrac{1}{2}E^2
%\\
&=& \dfrac{1}{2} e^{-2\alpha+4\beta} f^2 \dot{A}^2
 \,.
\end{eqnarray}
Moreover, the Einstein field equations are given by 
\begin{eqnarray}
\label{Fr1}
3 M_P^2 ~\big( \dot{\alpha}^2 - \dot{\beta}^2 \big) &=& V(\phi)+ \frac{1}{2}\dot{\phi}^2+\rho_E~,\\
\label{Fr2}
M_P^2 ~\big( \ddot{\alpha} + 3\dot{\alpha}^2 \big) &=& V(\phi)+\frac{1}{3}\rho_E~,\\
\label{ani}
3 M_P^2 ~\big(\ddot{\beta}+3\dot{\alpha}\dot{\beta}\big) &=& 2\rho_E \, .
\end{eqnarray}
The total energy density governing the dynamics of inflation is given by the right hand side of 
Eq. \eqref{Fr1} in  which the last term comes from the gauge field. In order to have a long period of slow-roll inflation, we require that 
\begin{eqnarray}
V(\phi) \gg \frac{1}{2}\dot{\phi}^2 \,,~ \rho_E  \,,~ M_P^2 \dot{\beta}^2 \, .
\label{condition_Inf}
\end{eqnarray}

In order to have small anisotropy, the electric field energy density should be small compared to the inflaton energy density $\rho_{\phi}$. Defining the parameter  
 $R_E$ as the fraction of the electric field energy density to the inflaton energy density,  
\begin{eqnarray}
R_E \equiv \dfrac{\rho_E}{\rho_{\phi}}
\simeq
 \dfrac{E^2}{2V}~,
\label{R_E}
\end{eqnarray}
we require that $R_E \ll 1$. 

As seen from Eq. \eqref{ani}, the fate of the anisotropy expansion rate $\dot{\beta}$ depends on the behavior of gauge field energy density. For a homogeneous gauge field satisfying the Maxwell equation \eqref{EoM-A}, the electric field energy density is given by
\begin{eqnarray}
\label{ani2}
\rho_E &=& \dfrac{1}{2} q_0^2 f^{-2} e^{-4\alpha-4\beta}~,
\end{eqnarray}
where $q_0$ is a constant of  integration. In the critical 
case with $f \propto e^{-2\alpha}$, and neglecting the anisotropy $\beta \ll 1$,  the electric field energy  density  remains constant and no anisotropy is produced during inflation. At the level of perturbations, this critical case corresponds to the situation where the gauge field perturbations become pure isocurvature, with no contribution to primordial power spectrum. 

More generally, choosing the conformal coupling in the form, 
\begin{equation}
f(\phi) = \exp\left[ \dfrac{2c}{M_P^2}\int\dfrac{V}{V_{,\phi}} \dd \phi \right] \propto  e^{-2 c \alpha} \, ,
\label{f_general2}
\end{equation}
with $c>1$ being a model parameter, it is shown in \cite{Watanabe:2009ct} that the system reaches the attractor regime in which the electric field energy density furnishes a small but a constant fraction of the total energy density. More specifically, it is shown in \cite{Watanabe:2009ct} that 
\begin{eqnarray}
R_E = \dfrac{1}{2}I \epsilon_H \,,~~~~~~~ \Sigma = \dfrac{2}{3} R_E \, ,
\label{R-Sigma}
\end{eqnarray}
where $I \equiv (c-1)/c$ and $\epsilon_H \equiv - \dfrac{\ddot{\alpha}}{\dot{\alpha}^2}$ is the slow-roll parameter. 

In this limit the kinetic coupling function can be obtained as~\cite{Watanabe:2009ct}
\begin{eqnarray}
&f(\phi)\propto e^{-2\alpha}\sqrt{1+\Omega\, e^{-4(c-1)\alpha}}\,,
\label{f_general}
\end{eqnarray}
where $\Omega$ is an integration constant. Furthermore,  the  background  electric field  energy density  as a function of time 
is given by~\cite{Watanabe:2009ct}
\begin{align}
\label{rho general}
\rho_E(t) = \frac{\rho_E^{\rm att}}{1+\Omega\, e^{-4(c-1)\alpha}} \, , \quad \quad  
\rho_E^{\rm att} \equiv \dfrac{3}{2} M_P^2 I \epsilon_H  H^2 \,.
%\\ &\rho_E^{\rm att} \equiv \dfrac{3}{2} M_P^2 I \epsilon_H  H^2 \,. \label{rho att}
\end{align}
Consequently $\Omega$ can be determined by the initial value of $\rho_E(t)$, e.g. consider $t_{\rm ini}$ when $\rho_E(t_{\rm ini})\equiv \rho_E^{\rm ini}$; then $\Omega$ is given by 
\begin{eqnarray}
\Omega=\frac{\rho_E^{\rm att}}{\rho_E^{\rm ini}}-1 \, .
\label{Omega2}
\end{eqnarray}
This can also be rewritten as
\begin{eqnarray}
\Omega=\left(\dfrac{E_{\rm att}}{E_{\rm ini}}\right)^2-1 \,,
\label{Omega}
\end{eqnarray}
where
\begin{eqnarray}
E_{\rm att}^2 \equiv  3 M_P^2 H^2 I \epsilon_H \,.
\label{E_att}
\end{eqnarray}

In general, when the system is not in its attractor regime, one can extend Eq. (\ref{R-Sigma}) to the following form 
\ba
\label{RE-general}
R_E(t) = \frac{1}{2} I \epsilon_H \left(\frac{E(t)}{E_{\rm att}} \right)^2 \, .
\ea

It was argued in \cite{Watanabe:2009ct} that  as the background expands the term containing $\Omega$ in the denominator of Eq. \eqref{rho general} falls off exponentially. Therefore, waiting long enough, one can neglect the term containing $\Omega$ in $\rho_E(t)$, yielding to the attractor solution $\rho_E(t) \rightarrow \rho_E^{\rm att}$.  Typically, one expects this happens after $1/(c-1)$ e-folds. However, as we shall see later,  from the observational constraint on the amplitude of quadrupole anisotropy one requires
that $I \sim (c-1)  \lesssim 10^{-7}$. This means that in order for the system to reach the attractor regime, one has to wait for  about $10^{7}$ e-folds. It was argued in \cite{Fujita:2017lfu} that during this long period, the stochastic effects of the gauge field perturbations will build up and prevent the system from reaching to its final attractor regime.  The  issue of whether or not the system will reach to its attractor phase and its observational implications were also studied in \cite{Naruko:2014bxa}.  One of the main goal of this paper is to study the roles of stochastic effects more systematically.

Next, we review the statistical properties of the curvature perturbations in this model. Since the dominant  contributions in statistical anisotropies come from the matter fields perturbations, one can neglect the gravitational back-reactions including the effects of 
$\beta$ in the metric \cite{Emami:2013bk}. First, we look at the evolution of the gauge field perturbation $\delta \boldsymbol{A}$ which, in Fourier space, can be expanded as
\begin{eqnarray}
\delta \boldsymbol{A}(\eta,\boldsymbol{x})&=& \sum_{\lambda = \pm} \int \frac{d^3k} {\left(2\pi\right)^3} \,  e^{i\boldsymbol{k}.\boldsymbol{x}} ~\boldsymbol{e}^\lambda(\hat{\boldsymbol{k}}) \left[ \delta A_\lambda(\eta,k)\,\hat{a}^\lambda_{\boldsymbol{k}} +  \delta A_{\lambda}^{*}(\eta,k)\, \hat{a}^{\lambda \dagger}_{-\boldsymbol{k}} \right] \,.
\end{eqnarray}
Here $\eta$ is the conformal time ${\rm d}\eta = e^{-\alpha} {\rm d}t$ and $\hat{a}^\lambda_{\boldsymbol{k}}$ and $\hat{a}^{\lambda \dagger}_{-\boldsymbol{k}}$ are the annihilation and creation operators, respectively, satisfying the commutation relation,
\begin{eqnarray}
[a^\lambda_{\boldsymbol{k}}, \ a^{\lambda' \dagger}_{-\boldsymbol{k}'}] &=& (2\pi)^3\delta^{\lambda \lambda'}\delta(\boldsymbol{k} + \boldsymbol{k}') \,.
\end{eqnarray}
Also $\boldsymbol{e}^\lambda(\hat{\boldsymbol{k}})$ represents the circular polarization vectors,  normalized such that
\begin{eqnarray}
\boldsymbol{e}^\lambda(\hat{\boldsymbol{k}}).\boldsymbol{e}^{\lambda'}(\hat{\boldsymbol{k}})&=&\delta^{\lambda\lambda'} \,,\\
\boldsymbol{\hat{k}}.\boldsymbol{e}^\lambda(\hat{\boldsymbol{k}}) &=& 0 \,,\\
i\hat{\boldsymbol{k}} \times \boldsymbol{e}^\lambda &=& \lambda \boldsymbol{e}^\lambda \,,\\
\boldsymbol{e}_\lambda(\hat{\boldsymbol{k}}) &=& \boldsymbol{e}^*_{\lambda}(-\hat{\boldsymbol{k}}) \,,\\
\sum_{\lambda = \pm} e_i^{\lambda}(\hat{\boldsymbol{k}})~e_j^{\lambda}(\hat{\boldsymbol{k}}) &=& \delta_{ij}-\hat{k}_i \hat{k}_j \,.
\label{ee}
\end{eqnarray}

Assuming $f \propto \eta^n$ with $n=2c$ from Eq. (\ref{f_general2}), and starting with the Bunch-Davies (Minkowski) initial condition, the  electric and magnetic components of the gauge field perturbations on super-horizon limit are given by
\cite{Bartolo:2012sd}
\begin{equation} \label{E_B}
\dfrac{\delta E_k}{H^2} \propto k^{\frac{1}{2}-n} ~ (-\eta)^{2-n} \ , \quad \dfrac{\delta B_k}{H^2} \propto k^{\frac{3}{2}-n} ~ (-\eta)^{3-n} \,.
\end{equation}

As we mentioned before, for the critical case with $n=2$ corresponding to $c=1$,  the electric field perturbations  are frozen on super-horizon scales, behaving as pure isocurvature perturbations.   In the general case where $f \propto e^{-2c\alpha}$ with $c>1$,  the electric field perturbations $\delta E_i$ contribute to the power spectrum  of curvature perturbation, $\zeta \equiv -H \delta \phi / \dot{\phi}$,  yielding the correction in power spectrum   \cite{ Watanabe:2010fh, Bartolo:2012sd, Emami:2013bk, Abolhasani:2013zya}, 
\begin{eqnarray}
{\cal P}_\zeta^E = 24 I N_k^2 {\cal P}_\zeta^{(0)} \left(\dfrac{E}{E_{\mathrm{att}}}\right)^2 (\hat{\bf k}.\hat{\bf p})^2 \,, 
\end{eqnarray}
where $\hat{\bf p}$ is the preferred anisotropic direction in the sky (the $x$-axis in our example), $N_k$ is the number of e-folds when the mode of interest $k$ leaves the horizon  
and ${\cal P}_\zeta^{(0)}$ represents the isotropic power spectrum, 
\begin{eqnarray}
{\cal P}_\zeta^{(0)} &=& \dfrac{H^2}{8\pi^2 M_P^2 \epsilon_H} \,.
\end{eqnarray}

Comparing the total power spectrum with the  quadrupolar anisotropic estimator $g_*$ defined by
\begin{eqnarray}
{\cal P}_\zeta \big(\boldsymbol{k}\big) = {\cal P}_\zeta^{(0)}(k) \Big(1+g_*  (\hat{\bf k}.\hat{\bf p})^2 \Big) \,,
\end{eqnarray}
we obtain 
\begin{eqnarray}
g_* = 24 I N_k^2 ~ \left(\dfrac{E}{E_{\mathrm{att}}}\right)^2 \,.
\label{g*}
\end{eqnarray}

To solve the flatness and the horizon problems, we need at least $N_{\rm CMB} \sim 60$ for the CMB scale modes. Also, from cosmological observations, we have  
$\abs{g_*} \lesssim 10^{-2}$ \cite{Ade:2015lrj,Kim:2013gka}.  Therefore, if the system had been in the attractor regime till $N_k = N_{\rm CMB}$, the background electric field energy density was given by its attractor value and we obtain the following constraint on anisotropy parameter, 
\begin{eqnarray}
I \lesssim 10^{-7} \left(\dfrac{N_{\rm CMB}}{60}\right)^2\,.
\label{I}
\end{eqnarray}
The problem with this small upper bound on $I$ is that it takes the system a long period of inflation to reach to its attractor regime. Specifically, as mentioned before, in order for the system to reach the attractor regime
one has to neglect the effects of the term containing the parameter $\Omega$ in the denominator of Eq. \eqref{rho general}. This is justified if one waits $1/(c-1) \sim 1/I \gtrsim 10^{7}$ e-folds of inflation. It is natural to expect that the stochastic effects of the gauge field perturbations may build up during this long period of inflation, spoiling the validity of the attractor assumption  \cite{Fujita:2017lfu}. We study this question more systematically in next Sections. 

%%%%%%%%%%%%%%%%%%%%%%%%%%%%%%%%%%%%%%%%%%%%%%%%%%%%%
\section{Stochastic Effects of Gauge Fields}
\label{stoc-rev}

In this Section we study the stochastic effects of the gauge fields on the  would-be attractor solution of \cite{Watanabe:2009ct}.

The profile of the electric and the magnetic perturbations on superhorizon scales are given in  Eq. \eqref{E_B}. As seen,  with $n  \gtrsim 2~ (c \gtrsim 1)$,  the electric field does not decay on superhorizon scales while  the magnetic field falls off exponentially so we study  the electric field perturbations. 

The equation of motion for the electric field components $E_i$ is given by 
\begin{eqnarray}
\ddot{E}_i - e^{-2\alpha}\nabla^2 E_i + 5H \dot{E}_i  + \left[ 6H^2 \left(1-\dfrac{1}{3}\epsilon_H\right) + \frac{\ddot{f} + H\dot{f}}{f} - 2\frac{\dot{f}^2}{f^2}\right]E_i
=0 \,. \label{E_EoM}
\end{eqnarray}
To employ the stochastic formalism, we decompose the fields and their conjugate  momenta into the UV 
and the IR parts. We are interested in the dynamics of the IR modes which are affected by the UV modes. The quantum fluctuations of the UV modes are represented by a Gaussian white noise, of which its amplitude is determined by the Hubble expansion rate $H$ during inflation.

With this discussion in mind, the  electric field and its momentum are decomposed in IR and UV modes as follows:
\begin{eqnarray}
	\boldsymbol{E}(t,\boldsymbol{x}) &=& \boldsymbol{E}^{\rm IR}(t,\boldsymbol{x}) + \sqrt{\hbar} \,~ \boldsymbol{E}^{\rm UV}(t,\boldsymbol{x}) \,,\label{E-UV-IR-dec}\\
	\boldsymbol{\dot{E}} (t,\boldsymbol{x}) &=& \boldsymbol{\Pi}_{\boldsymbol{E}}^{\rm IR}(t,\boldsymbol{x}) + \sqrt{\hbar}\,~ \boldsymbol{\Pi}_{\boldsymbol{E}}^{\rm UV}(t,\boldsymbol{x}) \,.
	\label{EPi-UV-IR-dec}
\end{eqnarray}
The UV parts are accompanied by a factor $\sqrt{\hbar}$ to exhibit the quantum nature of the short modes. 
The above decomposition is usually performed through an appropriate window function. Here we employ the Heaviside function  $ \Theta \left(k-\varepsilon aH\right)$  to separate the  long and short modes 
 in which $\varepsilon$ is a  dimensionless parameter satisfying $\varepsilon \ll 1$.  Putting these together, we have 
\begin{eqnarray}
\boldsymbol{E}^{\rm UV}(t,\boldsymbol{x}) &=& \int \frac{d^3k} {\left(2\pi\right)^3} \, \Theta \left(k-\varepsilon aH\right) \boldsymbol{E}(t,\boldsymbol{k}) e^{i\boldsymbol{k}.\boldsymbol{x}} \,,\label{E-UV}\\
\boldsymbol{\Pi}_{\boldsymbol{E}}^{\rm UV}(t,\boldsymbol{x}) &=& \int \frac{d^3k} {\left(2\pi\right)^3} \, \Theta \left(k-\varepsilon aH\right) \boldsymbol{\dot{E}}(t,\boldsymbol{k}) e^{i\boldsymbol{k}.\boldsymbol{x}} \,,\label{EPi-UV}\\
\boldsymbol{E}(t,{\boldsymbol{k}}) &=& \sum_{\lambda = \pm} \boldsymbol{e}^\lambda(\hat{\boldsymbol{k}}) \left[ E_\lambda(t,k)\,\hat{a}^\lambda_{\boldsymbol{k}} +  E_{\lambda}^{*}(t,k)\, \hat{a}^{\lambda \dagger}_{-\boldsymbol{k}} \right] \,.
\label{E_lambda}
\end{eqnarray}
Here $E_\lambda(t,k)$ is the mode function which satisfies the following equation,  
\begin{eqnarray}
E''_\lambda - \dfrac{4}{\eta} E'_\lambda  + \left[ k^2 -\dfrac{(n-2)(n+3)}{\eta^2} \right]E_\lambda
&=& 0 \,,\label{E_short2}
\end{eqnarray}
where  we have assumed $f \propto \eta^n$. 

Imposing the Bunch-Davies (Minkowski) initial condition  for the canonically normalized field $ \eta^{-2}E_{\lambda} $ when the modes are deep inside the horizon $( k\eta \rightarrow -\infty )$ , we obtain the following solution for the electric field mode function, 
\begin{eqnarray}
E_{\lambda} = - i \dfrac{\sqrt{\pi}}{2}~k H^2~\eta^{5/2} ~H^{(1)}_{n+\frac{1}{2}}(-k\eta) \,.
\end{eqnarray}
Using the small argument limit of the Hankel function, 
%superhorizon approximation $k\eta \rightarrow 0$ of the Hankel function,
\begin{eqnarray}
H^{(1)}_{n + \frac{1}{2}}(-k\eta) \rightarrow  -i \dfrac{(n - {1 \over 2})!}{\pi}\bigg({2 \over -k\eta}\bigg)^{n + \frac{1}{2}}\, \quad \quad  (k\eta \rightarrow 0) , 
\label{Hanckel_limit}
\end{eqnarray}
we see that for $n=2$ the electric field has the near scale invariant amplitude  on superhorizon scales  $k\eta \rightarrow 0$,   
\begin{eqnarray}
E_\lambda \simeq
{3 H^2 \over \sqrt{2}k^{3/2}}  \quad \quad  (k\eta \rightarrow 0)   \, .
\label{E_superhorizon}   
\end{eqnarray}

\begin{figure}[t]
		\includegraphics[width=\linewidth]{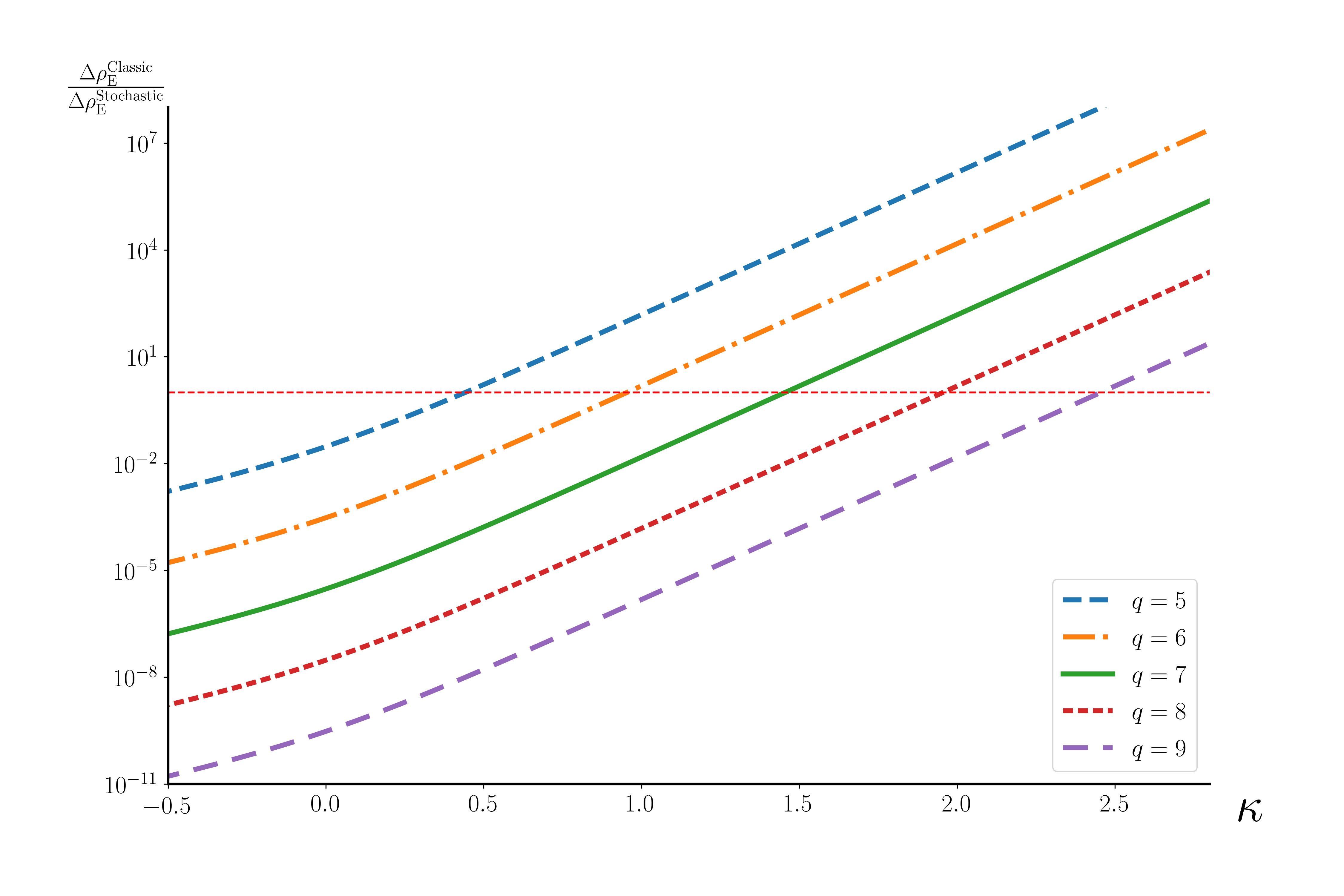}
		\caption{The ratio of the classical electric energy density to the stochastic one  as a function of the initial condition parameter $\kappa$  defined in Eq.~\eqref{q} for different values of $q$. The dashed line is related to the balance point; $\Delta \rho_E^{\rm classical}=\Delta \rho_E^{\rm stochastic}$.}
		\label{fig:ClastoSto}
	\end{figure}

Before continuing, it is instructive to compare  the stochastic evolution of the electric field energy density with the classical one. The stochastic kicks from the vacuum fluctuations  on the electric field energy density during one e-fold can be estimated as, 
\begin{eqnarray}
\Delta \rho_E^{\rm stochastic} \simeq{\cal P}_E \,,
\end{eqnarray}
where ${\cal P}_E$ is the dimensionless power spectrum of the electric field  fluctuations in the super-horizon 
limit,
\begin{eqnarray}
{\cal P}_E = \dfrac{k^3}{2\pi^2}|E_\lambda|^2=\dfrac{9H^4}{4\pi^2} \,.
\end{eqnarray}
On the other hand,  the classical evolution of $\rho_E$  during one e-fold  can be estimated from Eq. \eqref{rho general} as
\begin{eqnarray}
\Delta \rho_E^{\rm classical} \simeq \dfrac{2 \Omega (c-1)}{(1+\Omega)^2}E_{\rm att}^2+{\cal O}(c-1)^2 \,.
\end{eqnarray}
Hence the ratio of the classical evolution to the stochastic one is given by
\begin{eqnarray}
\dfrac{\Delta \rho_E^{\rm classical}}{\Delta \rho_E^{\rm stochastic}} \simeq \dfrac{1.5}{c} \times 10^{8+2(\kappa-q)}(1+10^{2 \kappa}) \, ,
\end{eqnarray}
where we have used  ${\cal P}^{(0)}_{\zeta} \simeq 2.2 \times 10^{-9}$ and defined the following dimensionless  parameters $q>0$ and $\kappa$, 
%\begin{eqnarray} c-1 &\equiv& 10^{-q} ~~~~~~~q>0\,, \label{q} \\ \left|\left(\dfrac{E_{\rm ini}}{E_{\rm att}}\right)^2 -1 \right| &\equiv& 10^{2\kappa} ~~~~~~~\kappa>-\infty\,. \label{r} \end{eqnarray}
\ba
\label{q}
q \equiv - \log_{10} (c-1) \,,  \quad  \quad 
2\kappa \equiv  \log_{10}  \Big| \big(\dfrac{E_{\rm ini}}{E_{\rm att}} \big)^2 -1 \Big| \, .
\ea
Note that from the constraint on $g_*$,  Eq. (\ref{I}), we expect typically that $q \gtrsim 7$. 
 As seen in Fig.\ref{fig:ClastoSto}, if $q=7$,  the stochastic kicks can not be ignored 
 for $\kappa \lesssim 1.5$. However,  for $\kappa>1.5$ the classical motion dominates and the system can reach to its classical attractor solution during one  e-folding time. However, as we shall show below, 
 for a general value of $\kappa$ the accumulative effects of the stochastic noises can be relevant if an exponentially large number of e-folds has passed.
%%%%%%%%%%%%%%%%%%%%%%%%%%%%%%%%%%%%%%%%%%%%%%%%%%%%%%%%%%%%%%%%%%%%%%%%%%%%%%%%%%%%%%%%%%%%%%%%%%%%%%%%%%%%%%%%%%%%%%%%%%%%%%%%%%%%%%%%%%%%%%%%%%
%%%%%%%%%%%%%%%%%%%%%%%%%%%%%%%%%%%%%%%%%%%%%%%%%%%%%%%%%%%%%%%%%%%%%%%%%%%%%%%%%%%%%%%%%%%%%%%%%%%%%%%%%%%%%%%%%%%%%%%%%%%%%%%%%%%%%%%%%%%%%%%%%%

To consider the stochastic effects more systematically, following the logic of \cite{Sasaki:1987gy, Nambu:1988je, Nakao:1988yi}, we expand Eq. \eqref{E_EoM} around $\boldsymbol{E}^{\rm IR}$ and $\boldsymbol{\Pi}_{\boldsymbol{E}}^{\rm IR}$ and keep the terms up to first order of $\sqrt{\hbar}$, obtaining  
\begin{eqnarray}
	\boldsymbol{\dot{\Pi}}_{\boldsymbol{E}}^{\rm IR} &=&  \sqrt{\hbar}\,\boldsymbol{\tau^E} + \dfrac{\nabla^2}{e^{2\alpha}} \boldsymbol{E}^{\rm IR} - 5H \boldsymbol{\Pi}_{\boldsymbol{E}}^{\rm IR}  - \left[ 6H^2\left(1-\dfrac{1}{3}\epsilon_H\right) + \frac{\ddot{f} + H\dot{f}}{f} - 2\frac{\dot{f}^2}{f^2}\right]\boldsymbol{E}^{\rm IR}
	 \,,
	\label{E_long}\\
	\boldsymbol{\dot{E}}^{\rm IR} &=& \boldsymbol{\Pi}_{\boldsymbol{E}}^{\rm IR} + \sqrt{\hbar}\,~ \boldsymbol{\sigma^E}  \,, \label{Edot_long2}
\end{eqnarray}
in which $\boldsymbol{\tau^E}$ and $\boldsymbol{\sigma^E}$ are the quantum noises, given by
\begin{eqnarray}
	\boldsymbol{\tau^E}(t,\boldsymbol{x}) = \varepsilon a H^2 \int \frac{d^3k} {\left(2\pi\right)^3} \, \delta \left(k-\varepsilon aH\right) \boldsymbol{\dot{E}}(t,\boldsymbol{k}) e^{i\boldsymbol{k}.\boldsymbol{x}} \,,\label{tau_Edot}\\
	\boldsymbol{\sigma^E}(t,\boldsymbol{x}) = \varepsilon a H^2 \int \frac{d^3k} {\left(2\pi\right)^3} \, \delta \left(k-\varepsilon aH\right) \boldsymbol{E}(t,\boldsymbol{k}) e^{i\boldsymbol{k}.\boldsymbol{x}} \,.\label{sigma_E}
\end{eqnarray}
In this view it is understood that  the noise terms $\boldsymbol{\tau^E}$ and $\boldsymbol{\sigma^E}$ behave as source terms originated  from short modes which affect  the evolution of long modes $\boldsymbol{E}^{\rm IR}$. The  correlation functions of  $\boldsymbol{\tau^E}$ and $\boldsymbol{\sigma^E}$ 
  in the Bunch-Davies vacuum $|0 \rangle$ are given by (see Appendix \ref{Noise} for more details);
\begin{eqnarray}
	\left \langle0 \left| \sigma^E_i (x_1) ~ \sigma^E_j (x_2) \right| 0\right\rangle 
	&\simeq&  {3 H^5 \over 2 \pi^2} ~j_0\big(\varepsilon aH | \bfx_1- \bfx_2 |\big)~ \delta_{ij}~ \delta\left(t_1-t_2\right) \,,\\
	\left \langle0 \left| \tau^E_i (x_1) ~ \tau^E_j (x_2) \right| 0\right\rangle 
	&\simeq&  \varepsilon^4   \dfrac{H^7}{6\pi^2} ~j_0\big(\varepsilon aH | \bfx_1- \bfx_2 |\big)~ \delta_{ij}~ \delta\left(t_1-t_2\right) \,,\\
	\left \langle0 \left| \sigma^E_i (x_1) \tau^E_i (x_2) + \tau^E_j (x_2) \sigma^E_i (x_1) \right| 0\right\rangle 
	&\simeq&  \varepsilon^2   \dfrac{H^6}{2\pi^2}~ j_0\big(\varepsilon aH | \bfx_1- \bfx_2 |\big)~ \delta_{ij}~ \delta\left(t_1-t_2\right) \,,
\end{eqnarray}
in which $x_i=(t_i,\boldsymbol{x}_i)$ and $j_0$ is the zeroth order spherical Bessel function. In addition,  the commutation relations of $\sigma^E_i$ and $\tau^E_i$ are given by 
\begin{eqnarray}
\label{correlation}
\left[ \sigma^E_i (x_1),\sigma^E_j (x_2) \right] &=& \left[  \tau^E_i (x_1),\tau^E_j (x_2) \right]=0, \\
\label{correlation2}
\left[ \sigma^E_i (x_1),\tau^E_j (x_2) \right] &= & i \varepsilon^5\frac{H^6}{6\pi^2}~j_0\big( \varepsilon aH | \bfx_1-\bfx_2|\big)~\delta_{ij}~\delta\left(t_1-t_2\right).
\end{eqnarray}
As it can be seen from Eqs.  \eqref{correlation} and  \eqref{correlation2},  the quantum nature of $\sigma^E_i$ and $\tau^E_i$ disappears when $\varepsilon$ is chosen small enough so 
one can treat them as classical noises. Therefore, on large-scale limit,  we have (see App. \ref{Noise})  
\begin{eqnarray}
\label{sigma-sigma}
\left\langle \sigma^E_i(t)\right\rangle&=&0, \qquad
\left\langle \sigma^E_i(t) \sigma^E_j(t')\right\rangle
\simeq \frac{3H^6}{2\pi^2}~\delta_{ij}~\delta(N-N')\,,
\\
\tau^E_i &\sim& \mathcal{O}( \varepsilon^2 ) \,.
%\label{sigma-sigma}
\end{eqnarray}
Here we have replaced  the time variable with the number of e-folds, $\alpha=N$, via $\mathrm{d} N=H\mathrm{d}t$.  

One can combine Eqs. \eqref{E_long} and \eqref{Edot_long2} in the slow-roll approximation where $\boldsymbol{\dot{\Pi}}_{\boldsymbol{E}}^{\rm IR} \sim 0$ to obtain the corresponding Langevin equation for 
the superhorizon modes $\boldsymbol{E}^{\rm IR}$  (where $\varepsilon \rightarrow 0$ and $e^{-2N}\nabla^2 \boldsymbol{E}^{\rm IR} \sim 0$). For this purpose, let us first simplify  the term inside the 
big bracket in Eq. \eqref{E_long}.  Noting that we work in the limit where   $c-1 \lesssim 10^{-7}$ and  $\epsilon_H \ll 1$ we have 
\begin{eqnarray}
6H^2\left(1-\dfrac{1}{3}\epsilon_H\right) + \frac{\ddot{f} + H\dot{f}}{f} - 2\frac{\dot{f}^2}{f^2} = -5H^2b + \mathcal{O}(\epsilon_H) + \mathcal{O}(c-1)^2 \,, 
\end{eqnarray}
in which we have defined  the parameter $b$ via 
\ba
\label{b-def}
b \equiv \frac{2\Omega(c-1)}{(1+\Omega)} \, ,
\ea
  Note that the parameter $b$ depends on the initial conditions through the parameter $\Omega$. In addition, from the definition of $\Omega$ 
in Eq. (\ref{Omega2}) we have the lower bound  $\Omega > -1$. Also we work with $c  >1$, so the sign of
the parameter $b$ is the same as $\Omega$.

Using the above approximations,  Eqs. \eqref{E_long} and \eqref{Edot_long2} can be combined to obtain 
 the desired  Langevin equation
\begin{eqnarray}
	\label{E_EoM3}
	\mathrm{d}\boldsymbol{E}^{\rm IR} = b ~\boldsymbol{E}^{\rm IR} ~\mathrm{d}N +  \sqrt{6}\frac{H^2}{2\pi}~\mathrm{d}\boldsymbol{W} \,. 
\end{eqnarray}
The first term is called  the ``drift term'' while the second one  is the ``diffusion term''. Here $\textbf{W}$ is a three dimensional (3D) Wiener process \cite{Evans} associated with a 3D normalized white classical noise $\boldsymbol{\xi}(N)$ defined via 
\begin{eqnarray}
    \boldsymbol{\sigma^E} &\equiv& \sqrt{6}\frac{H^3}{2\pi}\,\boldsymbol{\xi} \,,
    \\
	\mathrm{d}\boldsymbol{W}(N) &\equiv & \boldsymbol{\xi}(N) ~\mathrm{d}N \, , 
\end{eqnarray}
satisfying, 
\ba
\left\langle \xi_i(t)\right\rangle &=& 0 \,,\\
	\left\langle \xi_i(N) ~\xi_j(N')\right\rangle &=& \delta_{ij}~\delta(N-N')\,.
\ea
By defining the following  dimensionless stochastic variables, 
\begin{eqnarray}
\label{cal_E}
\boldsymbol{\mathcal{E}} \equiv \dfrac{\boldsymbol{E}^{\rm IR}}{E_{\rm att}}~\,,
~~~~~~~~~~~~~~~~~~
D \equiv 2\sqrt{\dfrac{{\cal P}_\zeta^{(0)}}{I}} \,,
\end{eqnarray}
the Langevin equation \eqref{E_EoM3} can be cast into the form of a dimensionless stochastic differential equation
\begin{eqnarray}
\label{E_EoM-dimles}
\mathrm{d}\boldsymbol{\mathcal{E}}(N) = b ~\boldsymbol{\mathcal{E}} ~\mathrm{d}N +  D~\mathrm{d}\boldsymbol{W}(N) \, .
\end{eqnarray}
This is our master equation for the following analysis. 

The general solution of Eq. (\ref{E_EoM-dimles}) is given by 
\begin{eqnarray}
\boldsymbol{\mathcal{E}}(N) = \boldsymbol{\mathcal{E}}_{\rm ini} \,e^{b N} +De^{b N}\int_{0}^{N}e^{-b s} \mathrm{d}\boldsymbol{W}(s)\,, 
\label{E}
\end{eqnarray}
with  the initial condition $\boldsymbol{\mathcal{E}}_{\rm ini}=\boldsymbol{\mathcal{E}}(0)$. The first term above represents the classical behavior of the electric field in the absence of stochastic noises,  
\begin{eqnarray}
\boldsymbol{\mathcal{E}}_{\rm cl}(N) = \boldsymbol{\mathcal{E}}_{\rm ini} \,e^{b N} \, .
\label{E_class}
\end{eqnarray}
As can be seen,  the constant parameter $b$ plays  crucial roles in stochastic differential equation Eq. \eqref{E_EoM-dimles}. The definitions \eqref{Omega} and \eqref{cal_E} help us to obtain the dependency of $b$ on initial condition as
\begin{eqnarray}
b = 2(c-1)(1-\mathcal{E}_{\rm ini}^2)\,.
\end{eqnarray}
We study the role of the parameter $b$ in more details below. But before doing so, let us calculate the  expectation  values (or mean values) and the variances related to vector stochastic quantity $\boldsymbol{\mathcal{E}}(N)$ 
obtained in Eq. (\ref{E}). 

Using the following properties of the stochastic integrals \cite{Evans}
\begin{eqnarray}
\Big\langle  \int_{0}^{T} G(t) {\rm d}W(t)  \Big\rangle = 0 \, , \quad  \quad 
\Big\langle \Big[ \int_{0}^{T} G(t) {\rm d}W(t) \Big]^2 \Big\rangle = \Big\langle \int_{0}^{T} G^2 {\rm d}t \Big\rangle \,,
\end{eqnarray}
we obtain 
\begin{eqnarray}
\left\langle 	\mathcal{E}_i(N)  \right\rangle &=& 	\mathcal{E}_i^{\rm ini}~e^{b N} = \boldsymbol{\mathcal{E}}_{\rm cl}(N)  \,,
\label{E_i}
\\
\left\langle \mathcal{E}_i^2(N)  \right\rangle &=&  \left((\mathcal{E}^{\rm ini}_{i})^2+\dfrac{D^2}{2b}\right)  ~e^{2b N}-\dfrac{D^2}{2b} \,,\label{E2}
\\
\delta \mathcal{E}_i^2(N) &=& %\delta (E^{ref}_{i})^2 ~e^{2b N} 
- \dfrac{D^2}{2b}(1-e^{2b N}) \,,
\\
\langle \mathcal{E}^2(N) \rangle   %&=&  \mathcal{E}_{\rm ini}^2  ~e^{2b N}-\dfrac{3D^2}{2b}(1-e^{2b N}) 
%\\
&=& \left( \mathcal{E}_{\rm ini}^2 + \dfrac{3D^2}{2b} \right)~e^{2b N} - \dfrac{3D^2}{2b}
\,,
\label{E2-aver}
\\
\delta \mathcal{E}^2(N) &=& %\delta (E^{ref}_{i})^2 ~e^{2b N} 
- \dfrac{3D^2}{2b}(1-e^{2b N}) \,,
\label{d-E}
\end{eqnarray}
where the variance is defined via $\delta  \mathcal{E}^2 \equiv \langle  \mathcal{E}^2 \rangle - \langle  \mathcal{E} \rangle^2 $. We see that  the above quantities depend on the number of e-fold $N$ and on the initial conditions via ${\cal E}_{\rm ini}$ and $b$.

Based on the above results, let us compare the case of  the classical feedback mechanism with no quantum noises to the case when the stochastic effects are included. In the classical treatment where  the contributions of the UV modes in electric energy density are neglected, the attractor regime is when ${\cal E}$ approaches to unity. If ${\cal E}_{\rm ini} >1$, the back-reactions from electric field  (the last term in Eq.  \eqref{EoM-Phi} ) grows, reducing the inflaton velocity and suppressing ${\cal E}$. On the other hand,  if ${\cal E}_{\rm ini} < 1$, the inflaton velocity increases  ${\cal E}$ towards unity through the coupling $f(\phi)$. As a side remark,  we note that switching off the feedback mechanism by going to the limit $c \rightarrow 1$, ${\cal E}$ stays on its initial value and there is no force to drive it to the attractor value. 

%%%%%%%%%%%%%%%%%%%%%%%%%%%%%%%%%%%%%%%%%%%%%%%%%%%%%%%%%%%%%%%%%%%%%%%%%%%%%%%%%%%%%%%%%%%%%%%%%%%%%%%%%%%%%%%%%%%%%%%%%%%%%%%%%%%%%%%%%%%%%%%%%%
	\begin{figure}[t]
		\includegraphics[width=\linewidth]{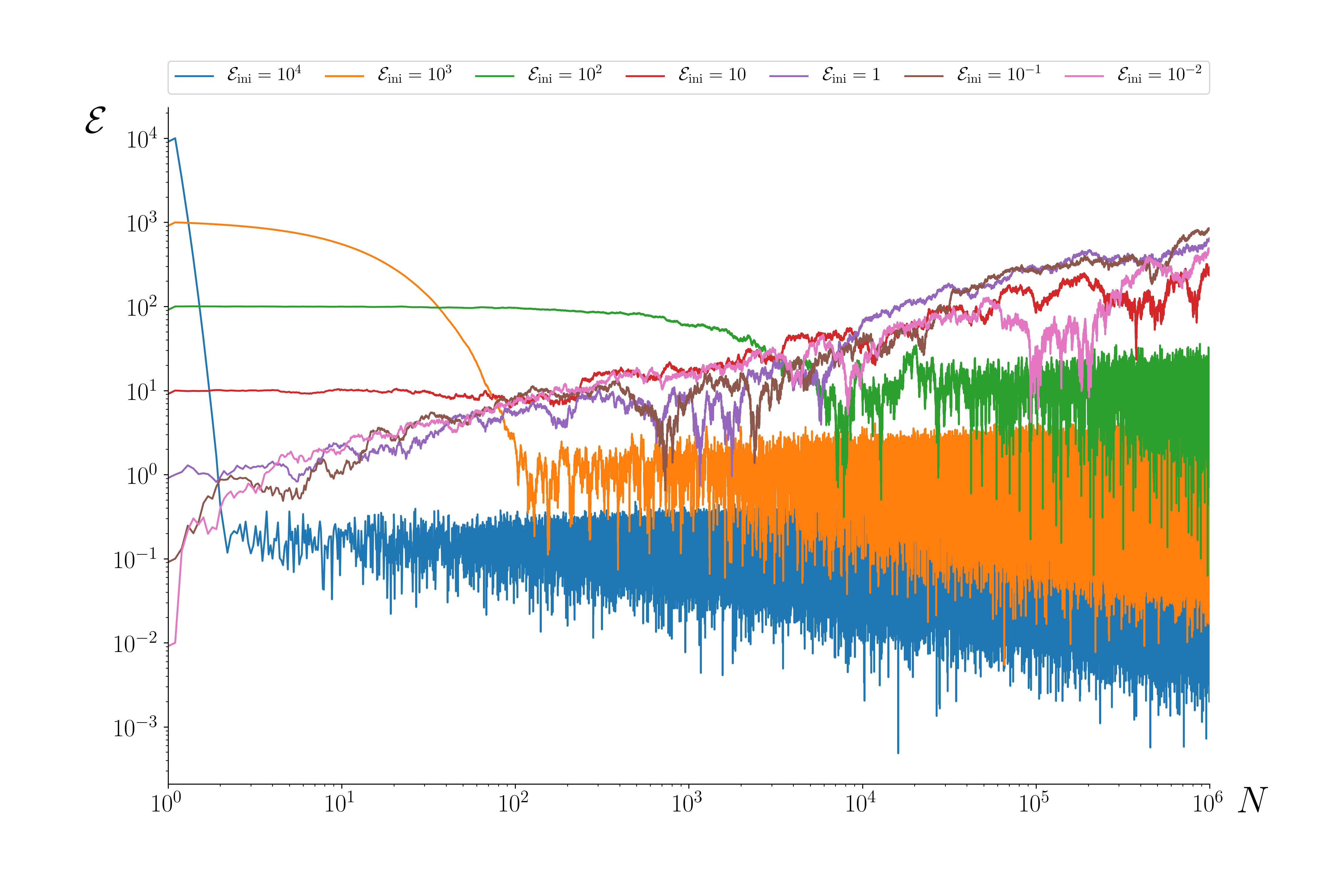}
		\caption{Evolution of $\mathcal{E}$ in terms of $N$ for different values of initial conditions ${\cal E}_{\rm ini}$. Cases with  $ {\cal E}_{\rm ini} >1 $ reach a  quasi-stable state limit while cases with
	${\cal E}_{\rm ini} \leq 1$  grow indefinitely.  The larger is ${\cal E}_{\rm ini} $, the faster system falls into its quasi-stable state. The enhanced oscillatory behaviour of the cases with  $ {\cal E}_{\rm ini} >1 $ is due to the mean-reverting process of an Ornstein-Uhlenbeck stochastic differential equation (see Sec. \ref{bmin}).
	}
		\label{fig:E-N}
	\end{figure}
%%%%%%%%%%%%%%%%%%%%%%%%%%%%%%%%%%%%%%%%%%%%%%%%%%%%%%%%%%%%%%%%%%%%%%%%%%%%%%%%%%%%%%%%%%%%%%%%%%%%%%%%%%%%%%%%%%%%%%%%%%%%%%%%%%%%%%%%%%%%%%%%%%

However the story is very different when the stochastic noises are switched on. In  Fig. \ref{fig:E-N} the 
numeric solution of Eq. \eqref{E_EoM-dimles} is plotted for different electric field initial conditions. As can be seen,  the fate of electric field is somewhat sensitive to its initial value. If $E_{\rm ini} > E_{\rm att}~ ( {\cal E}_{\rm in} >1$), the system falls  into a quasi-stable state (in the following we relate this quasi-stable state to the  stationary solution of probability density function of  electric  field). The larger is $E_{\rm ini} $, the faster the system falls into the quasi-stable state. On the other hand,  for $E_{\rm ini} \leq E_{\rm att} ~ ( {\cal E}_{\rm in} \leq1)$ the electric field energy density grows continuously and there is no attractor regime.  This is the non-trivial consequence of the stochastic effects of the gauge field perturbations.

\begin{figure}[t]
		\includegraphics[width=\linewidth]{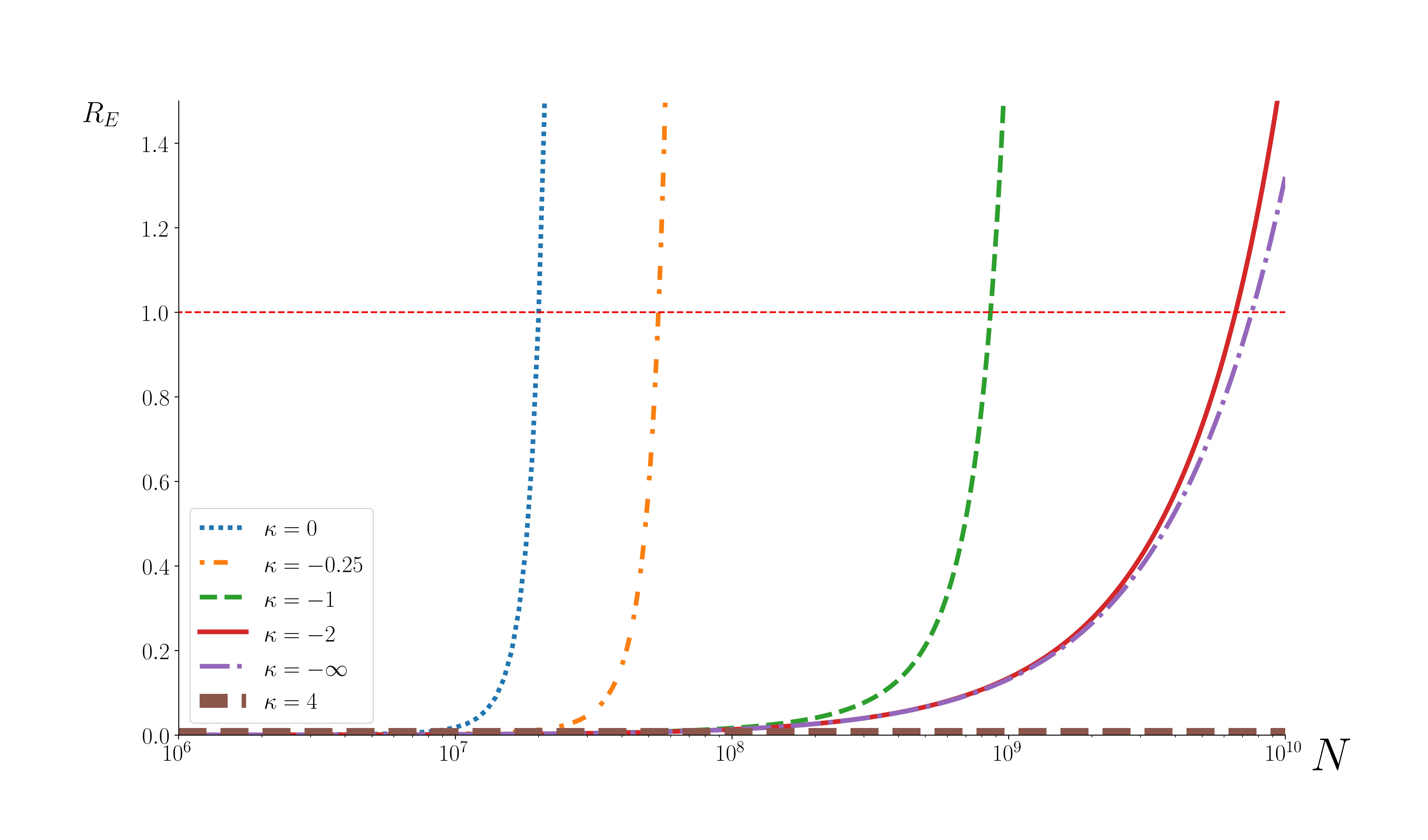}
		\caption{The behaviour of  $R_E$ as a function  of $N$ for $q=7$. As seen, for $\kappa>0$, the requirement $R_E \ll 1$ is never violated. For $\kappa<0$ this is violated for $N$ in the range of $10^7-10^{10}$.}
		\label{fig:RE-N}
	\end{figure}

One may worry that the accumulative effects of the stochastic noises may violate the condition  $R_E \ll1 $ required for a nearly isotropic background. Here we investigate this question while assuming that 
$\epsilon_H \simeq 10^{-2}$ and rewrite the parameter $\kappa$ in Eq. \eqref{q} for the IR part of dimensionless electric field as
\begin{eqnarray}
2\kappa \equiv  \log_{10}  \Big|{\cal E}^2_{\rm ini} -1 \Big| =  \log_{10}  \Big| \frac{b}{2 (c-1)} \Big|  \, .
\label{kappa}
\end{eqnarray}
For $\kappa \ge 0 ~ (\kappa <0)$, ${\cal E}_{\rm ini} \ge \sqrt 2  ~(< \sqrt 2 )$ and  
the initial electric field is somewhat larger (smaller)  than the classical attractor value. The limit $\kappa \rightarrow -\infty$ corresponds to the case where  the system starts with its attractor value, ${\cal E}_{\rm ini} =1,  b=0$,   in which  the Langevin equation \eqref{E_EoM-dimles} describes a pure Wiener process.  Fig \ref{fig:RE-N} presents the evolution of $R_E(N)$ for $q=7$. As can be seen, for $\kappa>0$, the nearly-isotropic slow-roll condition $R_E \ll 1$ is never violated. However, for $\kappa<0$,  the number of e-fold $N_{\rm vio}$ when $R_E$ approaches unity is  in the range of $10^7-10^{10}$. For a given value of $q$, we have $N_{\rm vio} \sim 10^{q}$. Incidentally, note that this is also  the number of e-folds required for the system to reach into its attractor limit  in the classical picture of \cite{Watanabe:2009ct}. As a result,  we conclude  that the stochastic effects do not allow for the system to reach into its classical attractor limit when $\kappa<0$, i.e. when $E_{\rm ini} \lesssim E_{\rm att}$.

%%%%%%%%%%%%%%%%%%%%%%%%%%%%%%%%%%%%%%%%%%%%%%%%%%%%%%%%%%%%%%%%%%%%%%%%%%%%%%%%%%%%%%%%%%%%%%%%%%%%%%%%%%%%%%%%%%%%%%%%%%%%%%%%%%%%%%%%%%%%%%%%%%
%%%%%%%%%%%%%%%%%%%%%%%%%%%%%%%%%%%%%%%%%%%%%%%%%%%%%%%%%%%%%%%%%%%%%%%%%%%%%%%%%%%%%%%%%%%%%%%%%%%%%%%%%%%%%%%%%%%%%%%%%%%%%%%%%%%%%%%%%%%%%%%%%%
Moreover, equations \eqref{E_i}-\eqref{d-E} tell us that the sign of $b$ is very important in determining 
the fate of the electric field and inflation dynamics. To describe the time evolution of the probability density function of ${\cal E} (N)$, we can employ the Fokker-Planck equation associated  to the Langevin equation \eqref{E_EoM3}. The Fokker-Planck equation for the probability density $f_{{\cal E}_i}(x,N)$ of the random variable ${\cal E}_i$ is given by 
\begin{eqnarray}
\dfrac{\partial f_{{\cal E}_i}(x,N)}{\partial N} = -b\dfrac{\partial}{\partial x} \bigg(x f_{{\cal E}_i}(x,N) \bigg) + \dfrac{D^2}{2} \dfrac{\partial^2}{\partial x^2}f_{{\cal E}_i}(x,N) \,.
\label{probab_E_i}
\end{eqnarray}
Intuitively, one can think of $f_{{\cal E}_i}(x,N) {\rm d}x$ as being the probability of ${\cal E}_i$ falling within the infinitesimal interval $[x,x+{\rm d}x]$. The existence of a stationary solution for the probability density of the electric field \eqref{E_EoM-dimles} depends on the sign of $b$, i.e. one  must compare the initial electric energy density with its attractor value. In the following we have classified the $f^2F^2$  model in three categories in the presence of stochastic noises.

%%%%%%%%%%%%%%%%%%%%%%%%%%%%%%%%%%%%%%%%%%%%%%%%%%%%%%
\subsection{$\mathbf{b>0} ~({\cal E}_{\rm ini} < 1)$}

If ${\cal E}_{\rm ini} <1$ (i.e. the initial energy density of the electric field is  smaller than its classical attractor value), the mean of the electric field energy density and its variance grow. 
At early stage when $bN \ll 1$, the mean  and the  variance grow linearly with time, 
\begin{eqnarray}
\langle \mathcal{E}^2 \rangle   &\approx& 
\left( 2b\mathcal{E}_{\rm ini}^2+3D^2 \right)N+\mathcal{E}_{\rm ini}^2 \,,
\label{E2-averb+}
\\
\delta \mathcal{E}^2 &\approx& 3D^2~N \,.
\label{d-Eb+}
\end{eqnarray}
Although the classical contribution \eqref{E_class} is initially under control by the feedback mechanism, but when $bN \gtrsim 1$, the stochastic noises grows exponentially and spoil the feedback mechanism. Moreover, for $b\geqslant0$, there is no stationary probability distribution and $R_E$ and $g_*$ grow linearly with time.

A spacial case is when ${\cal E}_{\rm ini}=0$ so there is no classical energy density. 
In this case   the system is described by a pure Brownian motion and 
$\langle \mathcal{E}^2 \rangle=\delta \mathcal{E}^2 = 3D^2 N$. The linear growth  of the variance with $N$
is the hallmark of the  Brownian motion. We see that  even with zero classical electric field energy density, the stochastic effects can grow and generate large  electric field energy density.

%%%%%%%%%%%%%%%%%%%%%%%%%%%%%%%%%%%%%%%%%%%%%%%%%%%%%%
\subsection{$\mathbf{b=0} ~({\cal E}_{\rm ini} \rightarrow 1)$}
If ${\cal E}_{\rm ini} = 1$,  the initial energy density of the electric field is equal to its classical  attractor 
value  and  the Langevin equation \eqref{E_EoM-dimles} describes a Wiener process with no drift, 
\begin{eqnarray}
\label{E_EoM-dimlesbzero}
\mathrm{d}\boldsymbol{\mathcal{E}} =   D~\mathrm{d}\boldsymbol{W} \,.
\end{eqnarray}
Correspondingly, the Fokker-Planck equation \eqref{probab_E_i} is simplified to 
\begin{eqnarray}
\dfrac{\partial f_{{\cal E}_i}(x,N)}{\partial N} =  \dfrac{D^2}{2} \dfrac{\partial^2}{\partial x^2}f_{{\cal E}_i}(x,N) \,,
\label{probab_E_i_b+}
\end{eqnarray}
which is the simplest form of a ``diffusion equation'' (also known as the heat equation). This partial differential equation, with the initial condition $f_{{\cal E}_i}(x,0)=\delta(x)$, has the solution
\begin{eqnarray}
\label{f_Ei}
f_{{\cal E}_i}(x,N) = \dfrac{1}{\sqrt{2\pi D^2 N}} \mathrm{exp}\Big({-\dfrac{x^2}{2D^2 N}}\Big)\,.
\end{eqnarray}
This shows that ${\cal E}_i$ has a Gaussian (normal) distribution,  denoted by   $\mathbb{N}(0,D^2N)$, describing a random walk process with zero mean  and with variance $D^2N$. 

The above density function allows us to compute the associated 
expectation  values as follows,  
\begin{eqnarray}
\left\langle 	\mathcal{E}(N)  \right\rangle &=& \int_{0}^{\infty}{\rm d}x ~x ~f_{{\cal E}}(x,N) 	=D \sqrt{8 N \over \pi} \,,
\label{E_b0}
\\
\langle \mathcal{E}^2(N) \rangle  
&=&  \int_{0}^{\infty}{\rm d}x ~x^2 ~f_{{\cal E}}(x,N) = 3D^2N \,,
\label{E-averb0}
\end{eqnarray}
where we have used the following form of the  probability density of ${\cal E} = (\sum_i { {\cal E}_i}^2)^{1/2}$ (see App. \ref{PDF} for more details),  
\begin{eqnarray}
\label{f_E_b0}
f_{{\cal E}}(x,N) &=& 2~ \sqrt{\dfrac{1}{2\pi D^6N^3}} ~x^2~  \mathrm{exp} \Big({-\dfrac{x^2}{2D^2 N}}\Big)\,.
\end{eqnarray}
Therefore $R_E$ and $g_*$ grow in time so the system does not 
reach to its attractor regime.

%%%%%%%%%%%%%%%%%%%%%%%%%%%%%%%%%%%%%%%%%%%%%%%%%%%%%%
\subsection{$\mathbf{b<0} ~({\cal E}_{\rm ini} > 1)$}
\label{bmin}

If ${\cal E}_{\rm ini}>1$ then  $E_{\rm ini} > E_{\rm att}$ and  Eq.  \eqref{E_EoM-dimles}
represents an Ornstein-Uhlenbeck (OU) stochastic differential
 equation~\cite{Evans}. The OU process $\boldsymbol{\mathcal{E}}$ $\left\{\boldsymbol{\cal E}_N, N\geq0\right\}$ is an example of a Gaussian process defined by Eq. \eqref{E_EoM-dimles} when $b<0$ and $D>0$ are constant parameters.  The random force $D \boldsymbol{\xi}$ is balanced by the frictional drift force $-|b| 	\boldsymbol{\mathcal{E}}$ and the process tends towards its long-term mean (mean-reverting process).
 This process admits a stationary probability distribution and has a bounded variance and a long-term mean. If the value of the field (process) is greater (less) than the mean value, then the drift will be negative (positive), e.g., the mean acts as an equilibrium level for the process. In this picture,  the classical feedback mechanism of \cite{Watanabe:2009ct} (responsible for the system to reach to its classical attractor regime)  is replaced by the mean-reverting process. This can be seen in  Fig.~\ref{fig:E-N}.

We note that while $D$ represents the amplitude of the diffusion  term, but $b$ represents the rate of the classical growth or the decay of the perturbations. For $b<0$ ($-1<\Omega<0$), we have a friction term which washes out the explicit dependence of the solution to the  initial conditions\footnote{ Here $b$ depends on the initial condition $\mathcal{E}_{\rm ini}$.}. As a result, for any initial condition $\mathcal{E}^{\rm ini}_{i}>1$, as $N \rightarrow \infty$,  we obtain
\begin{eqnarray}
	\label{limit_N_infinity}
	\left\lbrace \begin{array}{lc}
		\left\langle \mathcal{E}_i  \right\rangle \rightarrow 0 \,, \\
		\delta \mathcal{E}_i^2 \rightarrow  \dfrac{D^2}{2|b|} . \\
	\end{array}\right.
\end{eqnarray}
We  see that the distribution of $\mathcal{E}_i$ approaches $\mathbb{N}\Big(0,\dfrac{D^2}{2|b|}\Big)$ as $N \rightarrow \infty$ i. e.,  the solution after a long time settles down into a Gaussian distribution whose variance is $\dfrac{D^2}{2|b|}$.

The stationary solution (equilibrium state) of Eq. \eqref{probab_E_i}, $\partial f^{\rm eq}_{{\cal E}_i}/\partial N=0$, for $b<0$ is given by
\begin{eqnarray}
\label{f_Ei_station}
f^{\rm eq}_{{\cal E}_i}(x) = \sqrt{\dfrac{|b|}{\pi D^2 }} \mathrm{exp} \Big({-\dfrac{|b|}{D^2}}x^2 \Big)\,.
\end{eqnarray}
Correspondingly,  $f^{\rm eq}_{{\cal E}}(x)$,  the probability density functions of ${\cal E}$ 
(see App.\ref{PDF} for more details) is obtained to be
\begin{eqnarray}
\label{f_E_station}
f^{\rm eq}_{{\cal E}}(x) &=& 4~ \sqrt{\dfrac{|b|^3}{\pi D^6}} ~x^2~   \mathrm{exp} \Big({-\dfrac{|b|}{D^2}}x^2\Big)\, .
\end{eqnarray}
With the above density function, one can calculate various expectation values associated with ${\cal E}$ as follows:
\begin{eqnarray}
\left\langle \mathcal{E}  \right\rangle_{\rm eq} &=& \int_{0}^{\infty} {\rm d}x~x~f^{\rm eq}_{{\cal E}}(x)=\dfrac{2D}{\sqrt{\pi |b|}}\,,
\label{E_bmin}
\\
\langle \mathcal{E}^2 \rangle_{\rm eq} &=& \int_{0}^{\infty} {\rm d}x~x^2~f^{\rm eq}_{{\cal E}}(x)=\dfrac{3D^2}{2|b|} \,.
\label{E-averbmin}
\end{eqnarray}
Note that Eq.  \eqref{E-averbmin} is in agreement with Eq. \eqref{E2-aver} when $b<0$ and $N  \longrightarrow  \infty$.

It is interesting to examine  what happens if the gauge field is in the stationary state before the CMB scale modes leave the horizon.  In this case the amplitude of the electric field ${\cal E}$ is averaged over our observable universe  with the  distribution given in  Eq.  \eqref{f_E_station} at $N=N_{\rm CMB}$. Also one can estimate the equilibrium time when the system reaches to the stationary state and check the condition $R_E \ll 1$. In the following we  investigate the above questions and also  study  the classical attractor and the probability distribution of anisotropy in the equilibrium state.

\begin{figure}[t]
	\includegraphics[width=\linewidth]{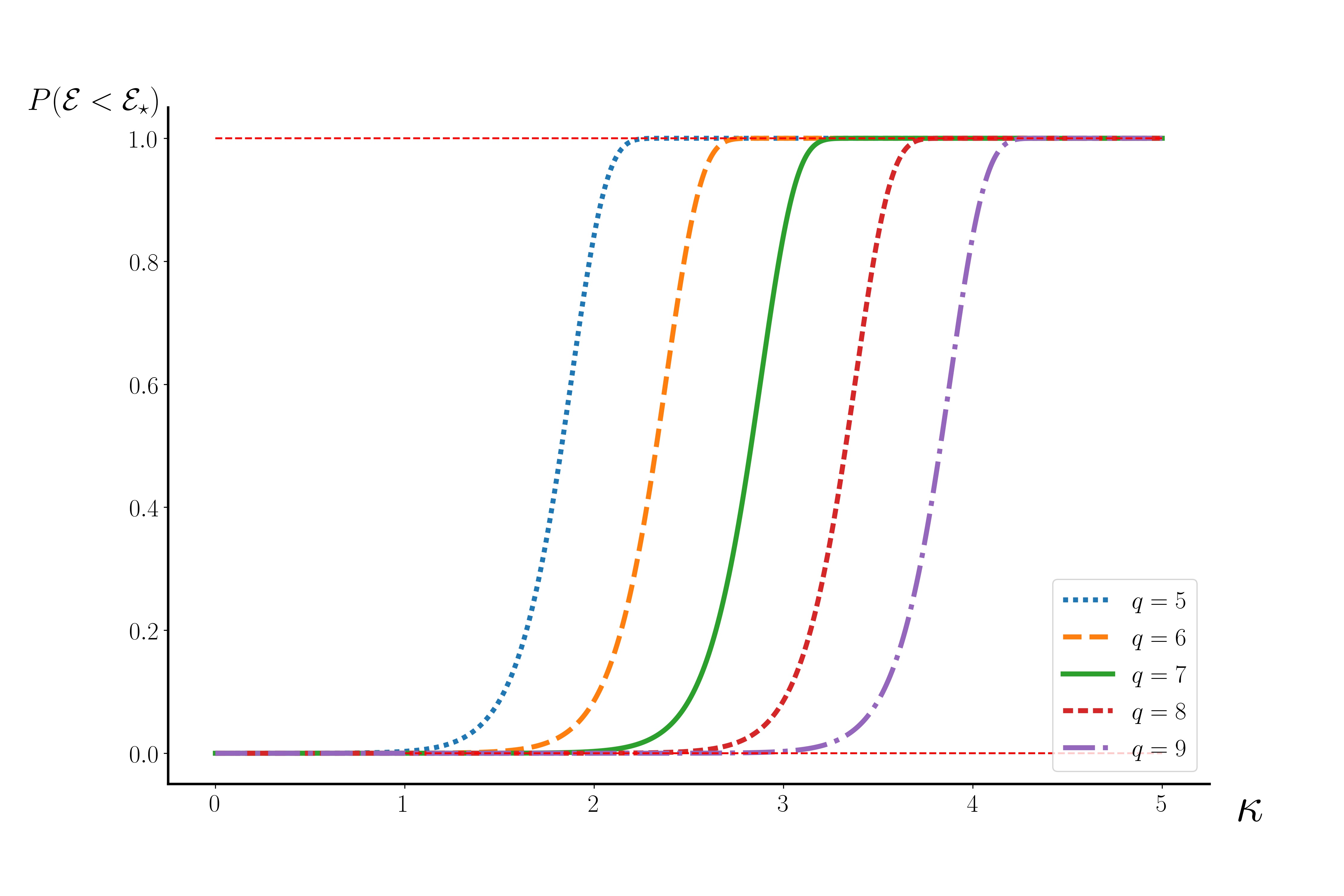}
	\caption{The probability of having $|g_*| < 10^{-2}$ for different values of the anisotropy  parameters $q$  in terms of the initial condition parameter $\kappa$. As can be seen, if the initial electric field  is high enough, one can obtain a probability of near unity  independent of  the value of $q$ .}
	\label{fig:P-r}
\end{figure}

%%%%%%%%%%%%%%%%%%%%%%%%%%%%%%%%%%%%%%%%%%%%%%%%%%%
\begin{itemize}
\item \textbf{Equilibrium time:}

First, let us  see when the system reaches to the equilibrium state. Let us define  $N_{\rm eq}$ as the time when 
$\langle \mathcal{E}^2(N_{\rm eq}) \rangle \rightarrow \langle \mathcal{E}^2 \rangle_{\rm eq}$. Formally, $N_{\rm eq} \rightarrow \infty$, but for practical purposes we can consider $N_{\rm eq}$ as when the ratio $\left( \langle \mathcal{E}^2(N_{\rm eq}) \rangle -\langle \mathcal{E}^2 \rangle_{\rm eq} \right)/\langle \mathcal{E}^2 \rangle_{\rm eq}$ drops to a small value say $10^{-2}$. 
With this approximation, and using  Eqs.~\eqref{E2-aver} and \eqref{E-averbmin}, 
we obtain  $N_{\rm eq} \simeq 10^{q-2\kappa}$. 
In particular, for $q=7$ one obtains $N_{\rm eq} \simeq 4.5 \times 10^7,~4.5 \times 10^4,~ 45 $ and $ 0.45$ for $\kappa=0,~1.5,~3$ and $4$ respectively. Comparing to the classical picture in which one has to wait for about $10^q$ e-folds in order for the system to reach to its attractor phase,  the stochastic effects can take the system to the equilibrium state after $10^{q-2\kappa}$ e-folds. The larger is $\kappa$, the faster the system falls into the quasi-stable state. This conclusion is supported in   Fig.~\ref{fig:E-N}.

\iffalse
%%%%%%%%%%%%%%%%%%%%%%%%%%%%%%%%%%%%%%%%%%%%%%%%%%%%%%%%%%%%%%%%%%%%%%%%%%%%%%%%%%%%%%%%%%%%%%%%%%%%%%%%%%%%%%%%%%%%%%%%%%%%%%%%%%%%%%%%%%%%%%%%%%
	\begin{figure}[t]
		\includegraphics[width=\linewidth]{Neqs40.jpg}
		\caption{Equilibrium time in terms of initial conditions for different model parameters $q$ when $s=40$.}
		\label{fig:Neq}
	\end{figure}
%%%%%%%%%%%%%%%%%%%%%%%%%%%%%%%%%%%%%%%%%%%%%%%%%%%%%%%%%%%%%%%%%%%%%%%%%%%%%%%%%%%%%%%%%%%%%%%%%%%%%%%%%%%%%%%%%%%%%%%%%%%%%%%%%%%%%%%%%%%%%%%%%%
\fi

\item \textbf{Isotropic condition:}

Second, we check the the small anisotropy  condition $R_E \ll 1$. %\textcolor{red}{As seen from Fig. \ref{fig:RE-N} if $\kappa>0$ the nearly-isotropic condition $R_E<1$ is never violated.}
Using the equilibrium state of $\langle \mathcal{E}^2 \rangle_{\rm eq} $ given by Eq. \eqref{E-averbmin},  and taking $\epsilon_H \sim 10^{-2}$, we obtain
\begin{eqnarray}
\label{RE-app}
R_E \sim  3.3 \times 10^{q-(11+2\kappa)} \, .
\end{eqnarray}

Hence the condition $R_E < 1$  is translated into $2\kappa > q-11$ which holds for a wide range of our parameterization ($q\leq11$). For example, with $q= 7$ and $\kappa =0.5$, we have $R_E \sim 10^{-5}$.

\item \textbf{Classical attractor:}

Third, let us  seek the condition under which the equilibrium state coincides with the classical attractor solution \cite{Watanabe:2009ct}. Requesting $\langle {\cal E}^2 \rangle_{\rm eq}=1$ yields
\begin{eqnarray}
\kappa \simeq q-4.1 \,.
\end{eqnarray}
%where $z=\pi/4,~2/3$ for $\langle {\cal E} \rangle_{\rm eq}=1$ and $\langle {\cal E}^2 \rangle_{\rm eq}=1$ respectively. 
Hence for $q=7$ this is translated into $\kappa \simeq 3$ corresponding to  $\rho_{E}^{\rm ini} \simeq 10^6 \rho_{E}^{\rm att}$, in which the equilibrium state coincides with the classical attractor solution. This equilibrium is reached  after $N_{\rm eq} \simeq 45$, significantly much less  than the classical attractor $\sim 10^{7}$ e-folds required naively in  \cite{Watanabe:2009ct}.

\item \textbf{Probability distribution of anisotropy:}

Finally,  we estimate the amplitude of the quadrupolar anisotropy $g_*$ and the probability that it satisfies the observational constraint. Combining Eqs. (\ref{RE-app}), (\ref{RE-general}) and (\ref{g*}), we obtain    
\begin{eqnarray}
g_* \simeq 5.7 \times 10^{q-2\kappa-4}\Big(\dfrac{N_{\rm CMB}}{60}\Big)^2 \, .
\end{eqnarray}
As we concluded  before, for $q=7$, the observational constraint $g_* < 10^{-2}$ is satisfied for $\kappa\gtrsim3$.  However, equipped with  the density function Eq. \eqref{f_E_station},  it is better to study this issue using  the language of the probability theory. The probability of having $g_* < 10^{-2}$ can be identified with the probability of  ${\cal E}<{\cal E}_*$ where we have defined 
\begin{eqnarray}
{\cal E}_*^2 \equiv \dfrac{10^{-2}}{24IN_k^2}\,.
\end{eqnarray}
This probability is given by
\begin{eqnarray}
P\left({\cal E}<{\cal E}_*\right) = \int_{0}^{{\cal E}_\star} {\rm d}x ~ f^{\rm eq}_{\cal E}(x) = \mathrm{Erf}\left(\sqrt{y}\right) -2\sqrt{\dfrac{y}{\pi}}~ \mathrm{exp}\left(-y\right) \,.
\end{eqnarray}
Here $\mathrm{Erf}$ is the error function and $y\equiv \dfrac{|b|}{D^2}{\cal E}_*^2$, which in terms of our parameterization, is given by $y=2.6 \times 10^{2\kappa+1-q}$. 

As shown in Fig \ref{fig:P-r}, this probability depends on $q$ and on the initial conditions. For example if $q=7$ and $\kappa=3$, the  probability that $g_*$ is less than $10^{-2}$ is $84.2\%$ if the electric field is in its equilibrium state before the CMB scale modes leave the horizon. The important lesson is that the probabilistic interpretation, arising from the effects of the stochastic noises, allows us to relax significantly the observational upper bound on $q$ (or $I$) as compared to the bound obtained classically  in Eq. \eqref{I}. In conclusion, the  $f^2F^2$ anisotropic inflation model  can be consistent with observations  if we  start with an electric field energy density larger than the  classical attractor value, corresponding to ${b<0} ~({\cal E}_{\rm ini} > 1)$.
\end{itemize}
%%%%%%%%%%%%%%%%%%%%%%%%%%%%%%%%%%%%%%%%%%%%%%%%%%%%%%%%%%%%%%%%%%%%%%%%%%%%%%%%%%%%%%%%%%%%%%%%%%%%%%%%%%%%%%%%%%%%%%%%%%%%%%%%%%%%%%%%%%%%%%%%%%
%%%%%%%%%%%%%%%%%%%%%%%%%%%%%%%%%%%%%%%%%%%%%%%%%%%%%%%%%%%%%%%%%%%%%%%%%%%%%%%%%%%%%%%%%%%%%%%%%%%%%%%%%%%%%%%%%%%%%%%%%%%%%%%%%%%%%%%%%%%%%%%%%%

%%%%%%%%%%%%%%%%%%%%%%%%%%%%%%%%%%%%%%%%%%%%%%%%%%%%%%%%%%%%%%%%%%%%%%%%%%%%%%%%%%%%%%%%%%%%%%%%%%%%%%%%%%%%%%%%%%%%%%%%%%%%%%%%%%%%%%%%%%%%%%%%%%
%\iffalse \textbf{page 15 Evans: density function and distribution function} \fi

\section{Stochastic Dynamics of Scalar Field}
\label{Scal_Sto}

To complete our studies of the stochastic effects on the system, here we present the  Langevin equation associated with the inflaton field given by  the Klein-Gordon equation \eqref{EoM-Phi}.

Following the standard method \cite{Sasaki:1987gy, Nambu:1988je, Nakao:1988yi}, we split  the inflaton field $\phi$ and its conjugate momentum into the long and short modes as follows,
\begin{eqnarray}
\phi(t,\mathbf{x}) &=& \phi^{\rm IR}(t,\mathbf{x}) + \sqrt{\hbar} ~ \phi^{\rm UV}(t,\mathbf{x}) \,,\label{Phi-UV-IR-dec}\\
\dot{\phi} (t,\mathbf{x}) &=& \pi^{\rm IR}(t,\mathbf{x}) + \sqrt{\hbar} ~\pi^{\rm UV}(t,\mathbf{x})
\,,\label{PhiPi-UV-IR-dec}
\end{eqnarray}
where, as before,  the UV modes are  defined in Fourier space as
\begin{eqnarray}
\phi^{\rm UV}(t,\mathbf{x}) &=& \int \frac{d^3k} {\left(2\pi\right)^3} \, \Theta \left(k-\varepsilon aH\right) \phi_{\mathbf{k}}(t) e^{i\mathbf{k}.\mathbf{x}}\,,\\
\pi^{\rm UV}(t,\mathbf{x}) &=& \int \frac{d^3k} {\left(2\pi\right)^3} \, \Theta \left(k-\varepsilon aH\right) \dot{\phi}_{\mathbf{k}}(t) e^{i\mathbf{k}.\mathbf{x}} \, .
\end{eqnarray}
%%%%%%%%%%%%%%%%%%%%%%%%%%%%%%%%%%%%%%%%%%%%%%%%%%%%%%%%%%%%%%%%%%%%%%%%%%%
\iffalse
In this kind of decomposition, we use the Heaviside function $\Theta \left(k\right)$ and a small dimensionless parameter $\varepsilon \ll 1$. Then UV modes can be defined in Fourier space as
\begin{eqnarray}
\phi^{\rm UV}(t,\mathbf{x}) = \int \frac{d^3k} {\left(2\pi\right)^3} \, \Theta \left(k-\varepsilon aH\right) \phi_{\mathbf{k}}(t) e^{i\mathbf{k}.\mathbf{x}} \,,\label{Phi-UV}\\
\pi_\phi^{\rm UV}(t,\mathbf{x}) = \int \frac{d^3k} {\left(2\pi\right)^3} \, \Theta \left(k-\varepsilon aH\right) \dot{\phi}_{\mathbf{k}}(t) e^{i\mathbf{k}.\mathbf{x}} \,.\label{PhiPi-UV}
\end{eqnarray}
\fi
%%%%%%%%%%%%%%%%%%%%%%%%%%%%%%%%%%%%%%%%%%%%%%%%%%%%%%%%%%%%%%%%%%%%%%%%%%%
The next step is to expand the Klein-Gordon equation \eqref{EoM-Phi} around $\phi^{\rm IR}$ and $\pi^{\rm IR}$ up to first order in $\sqrt{\hbar}$.  Neglecting the magnetic field and using Eq. \eqref{f_general2}, we rewrite the last two terms of Eq. \eqref{EoM-Phi} as
\begin{eqnarray}
V_{,\phi}(\phi) &=& V_{,\phi}(\phi^{\rm IR}) + \sqrt{\hbar} ~\phi^{\rm UV} V_{,\phi \phi}(\phi^{\rm IR}) + \mathcal{O}(\sqrt{\hbar})^2 \,,\\
\dfrac{f_{,\phi}}{f} E^2 &=&
I~ V_{,\phi} ~\mathcal{E}^2 + \sqrt{\hbar} I \bigg( \mathcal{E}^2 V_{,\phi \phi} \phi^{\rm UV} + 2\dfrac{V_{,\phi}}{E_{\rm att}} \boldsymbol{\mathcal{E}}.\boldsymbol{E}^{\rm UV}\bigg) \,.
\end{eqnarray}
Therefore, for long wavelength mode  we obtain
\begin{eqnarray}
\dot{\pi}^{\rm IR} -e^{-2\alpha}\nabla^2\phi^{\rm IR} + 3 H \pi^{\rm IR} + V_{,\phi} (1 - I~ \mathcal{E}^2) - \sqrt{\hbar}\,\tau= 0 \,,\label{phi_long}\\
\pi^{\rm IR} -\dot{\phi}^{\rm IR} + \sqrt{\hbar} ~ \sigma = 0 \,,\label{phi_long2}
\end{eqnarray}
in which the noises are defined via
\begin{eqnarray}
\tau (t,\mathbf{x}) &\equiv& \varepsilon a H^2 \int \frac{d^3k} {\left(2\pi\right)^3} \, \delta \left(k-\varepsilon a H\right) \dot{\phi}_{\mathbf{k}}(t) e^{i\mathbf{k}.\mathbf{x}} \,,\label{tau_Phi}\\
\sigma (t,\mathbf{x}) &\equiv& \varepsilon a H^2 \int \frac{d^3k} {\left(2\pi\right)^3} \, \delta \left(k-\varepsilon a H\right) \phi_{\mathbf{k}}(t) e^{i\mathbf{k}.\mathbf{x}} \,.\label{sigma_Phi}
\end{eqnarray}
As usual,  the operator $\phi_{\bfk }(t)$ can be expanded in terms of the creation and annihilation operators, $\phi_\bfk=a_\bfk\varphi_k+a^\dagger_{-\bfk}\varphi_{-k}^*$, in which $\varphi_k$ is the positive frequency mode function satisfying the following equation
\begin{eqnarray}
\ddot{\varphi}_k + 3 H \dot{\varphi}_k + \bigg( \dfrac{k^2}{e^{2N}} + M^2 \bigg) \varphi_k = 0 \,,
\label{phi_short}
\end{eqnarray}
where we have used $\langle \boldsymbol{\mathcal{E}}.\boldsymbol{E}(t,\boldsymbol{k}) \rangle \propto \langle \cos \theta \rangle = 0$ and  $\langle \mathcal{E}^2 V_{,\phi \phi}\rangle = \langle \mathcal{E}^2 \rangle \langle V_{,\phi \phi}\rangle$.  
%\begin{eqnarray} \langle \boldsymbol{\mathcal{E}}.\boldsymbol{E}(t,\boldsymbol{k}) \rangle &\propto& \langle \cos \theta \rangle = 0 \,,\\ \langle \mathcal{E}^2 V_{,\phi \phi}\rangle &=& \langle \mathcal{E}^2 \rangle \langle V_{,\phi \phi}\rangle \,. \end{eqnarray}
Furthermore,  $M^2=\langle V_{,\phi\phi} \rangle \big(1 - I  \langle \mathcal{E}^2 \rangle \big)$ is the average effective mass of the long wavelength perturbations of the inflaton. We see that the inflaton's effective mass receives corrections from the electric field which is the main reason for the system to attain its classical attractor solution in the mechanism of \cite{Watanabe:2009ct}. 

In order to solve Eqs. \eqref{phi_long} and \eqref{phi_long2}, we have to investigate the  correlations of the noises $\sigma$ and $\tau$. Starting with the Bunch-Davies  (Minkowski) vacuum $|0\rangle$, we obtain
\cite{Nakao:1988yi, Sasaki:1987gy}
\begin{eqnarray}
\label{11}
& \left \langle 0\left|\sigma\left(\bfx_1\right)\sigma\left(\bfx_2\right)\right| 0\right\rangle 
\approx \varepsilon^{\frac{2M^2}{3H^2}}\frac{H^3}{4\pi^2}j_0\big(\varepsilon aH | \bfx_1- \bfx_2 |\big)\delta\left(t_1-t_2\right),\\
\label{12}
& \left \langle 0\left|\tau\left( \bfx_1\right)\tau\left( \bfx_2\right)\right| 0\right\rangle
\approx \varepsilon^{\frac{2M^2}{3H^2}}\big(\frac{M^2}{3H^2}+\varepsilon^2\big)^2\frac{H^5}{4\pi^2}j_0\big(\varepsilon a H | \bfx_1-\bfx_2 |\big)\delta\left(t_1-t_2\right),\\
\label{13}
& \langle 0\left|\sigma( \bfx_1)\tau( \bfx_2)+\tau( \bfx_2)\sigma( \bfx_1) | 0\right\rangle  
\approx -2 \varepsilon^{\frac{2M^2}{3H^2}} 
\big(\frac{M^2}{3H^2}+ \varepsilon^2\big)\frac{H^4}{4\pi^2}j_0\big( \varepsilon aH | \bfx_1-\bfx_2|\big)\delta (t_1-t_2), 
\end{eqnarray}
with the following commutation relations 
\begin{eqnarray}
\label{correlation22}
\left[\sigma\left(\bfx_1\right),\sigma\left(\bfx_2\right)\right] &=&\left[\tau\left(\bfx_1\right),\tau\left(\bfx_2\right)\right]=0, \\
\label{correlation222}
\left[\sigma\left( \bfx_1\right),\tau\left(\bfx_2\right)\right] &= & i \varepsilon^3\frac{H^4}{4\pi^2}j_0\big( \varepsilon a H | \bfx_1-\bfx_2|\big)\delta\left(t_1-t_2\right).
\end{eqnarray}
From Eqs. \eqref{11}-\eqref{correlation222}, we find that for the parameter $\varepsilon$ in the range \cite{Nakao:1988yi, Sasaki:1987gy}
\begin{eqnarray}
e^{\frac{-3H^2}{|M^2|}} \ll \varepsilon^2 \ll \frac{|M^2|}{3H^2}\, ,
\end{eqnarray}
not only the quantum nature of $\sigma$ and $\tau$ becomes negligible but also the $\varepsilon$-dependence disappears from \eqref{11}-\eqref{13}. Furthermore, we obtain
\begin{eqnarray}
\tau &\approx& \dfrac{-M^2}{3H} \sigma \,,\\
\left<0\left|\sigma\left(\bfx_1\right)\sigma\left(\bfx_2\right)\right|0\right> &\approx& \frac{H^3}{4\pi^2}\delta\left(t_1-t_2\right)=\frac{H^4}{4\pi^2}\delta\left(N_1-N_2\right)\,. \label{stoch}
\end{eqnarray}
 Hence Eqs. \eqref{phi_long} and \eqref{phi_long2} reduce to a set of coupled classical Langevin equations,
\begin{eqnarray}
\dot{\pi}^{\rm IR} &=& e^{-2\alpha}\nabla^2\phi^{\rm IR} - 3 H \pi^{\rm IR} - V_{,\phi} (1 - I~ \mathcal{E}^2) - \dfrac{M^2}{3H}\sigma \,,\label{phi_long12}\\
\dot{\phi}^{\rm IR} &=&\pi^{\rm IR} + \sigma \, . \label{phi_long22}
\end{eqnarray}

Considering the  long wavelength modes where $e^{-2\alpha}\nabla^2\phi^{\rm IR} \rightarrow 0$ and 
working  in the  slow-roll limit where $\dot{\pi}_\phi^{\rm IR} \sim 0$,  the above two  equations can be combined to yield  the following Langevin equation for $\phi^{\rm IR}$
\begin{eqnarray}
\dfrac{\mathrm{d}\phi^{\rm IR}}{\mathrm{d}N}  + \dfrac{V_{,\phi}}{3H^2} \bigg(1  - I~\mathcal{E}^2 \bigg)=  \bigg(1-\dfrac{M^2}{9H^2}\bigg)\dfrac{H}{2\pi} \, \xi(N) \,,\label{Phi_long4}
\end{eqnarray}
in which $\xi(N)$ is the normalized white classical noise related to $\sigma (N)$ via
\begin{eqnarray}
\sigma \equiv  \dfrac{H^2}{2\pi} \xi(N)\, ,
\end{eqnarray}
 satisfying 
\begin{eqnarray}
\label{xi-noise2}
\big \langle \xi\left(N\right)\big \rangle = 0 \, , \quad \quad 
\big \langle \xi\left(N\right)\xi\left(N'\right)\big \rangle =\delta\left(N-N'\right) \, .
\end{eqnarray}

\begin{figure}[t]
	\includegraphics[width=\linewidth]{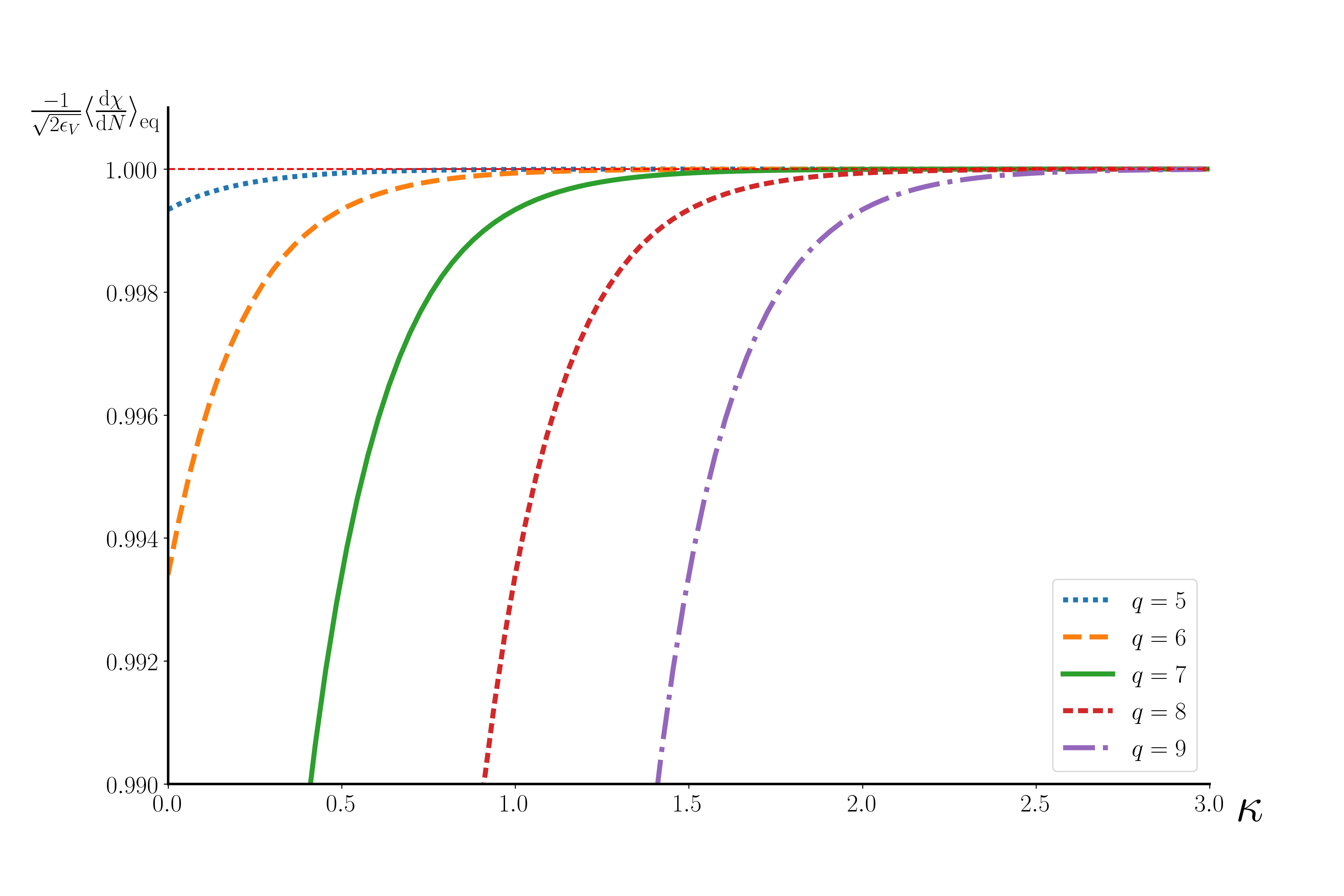}
	\caption{The terminal velocity of scalar field in terms of  $\kappa$ for various model parameters $q=6,7,8$.   As shown, a universal attractor regime for the scalar field is reached when the system is in stationary state where ${\cal E}_{\rm ini} \gg 1$ and $\kappa$ is large enough. }
	\label{fig:phi-dot}
\end{figure}

In the classical  attractor regime ${\cal E}=1$, and one recovers the known classical equation \cite{Watanabe:2009ct}
\begin{eqnarray}
\dfrac{\mathrm{d}\phi}{\mathrm{d}N}  + \dfrac{V_{,\phi}}{3H^2} \big(1  - I \big)= 0 \,.
\label{Phi_classic}
\end{eqnarray}
By defining the dimensionless field $\chi \equiv  \dfrac{\phi^{\rm IR}}{M_P}$, Eq. (\ref{Phi_long4}) is cast into
\begin{eqnarray}
\dfrac{\mathrm{d}\chi}{\mathrm{d}N}  + \sqrt{2\epsilon_V} \left(1  - I~\mathcal{E}^2 \right) =  \sqrt{2\epsilon_H {\cal P}_\zeta^{(0)}}~\bigg(1-\dfrac{M^2}{9H^2}\bigg) ~\xi(N) \,.
\end{eqnarray}
in which $\epsilon_V$ is the slow-roll parameter defined via $\epsilon_V  \equiv M_P^2 \left( V_\phi/V\right)^2/2$. Taking the expectation value of the above equation yields the average of the scalar field velocity,  
\begin{eqnarray}
\big \langle \dfrac{\mathrm{d}\chi}{\mathrm{d}N} \big \rangle = -\sqrt{2\epsilon_V} \left(1  - I~\big \langle \mathcal{E}^2 \big \rangle \right) \,.
\end{eqnarray}
In the absence of stochastic effects and for the classical attractor regime ($\mathcal{E}=1$), the velocity is slowed down by the factor $1/c$ \cite{Watanabe:2009ct}. Now switching on the stochastic effects 
 the velocity can be time-dependent based on the behaviour of $\langle \mathcal{E}^2 \big \rangle$, 
 see Eqs. \eqref{E2-averb+} and \eqref{E-averb0}. In the case with ${\cal E}_{\rm ini} \leq 1$ where there is 
 no stationary electric field probability distribution function, this may spoil inflation. However, in the equilibrium state  with ${\cal E}_{\rm ini} > 1$ and $b<0$,  we can obtain a terminal velocity and the system reaches to an attractor regime. 
 %(similar to the case of classical attractor regime in the absence of the stochastic noises)
The terminal velocity can be obtained from \eqref{E-averbmin} as
\begin{eqnarray}
\big \langle \dfrac{\mathrm{d}\chi}{\mathrm{d}N} \big \rangle_{\rm eq} = -\sqrt{2\epsilon_V} \left(1-\dfrac{3ID^2}{2|b|} \right) \,  ,  \quad \quad \big ( b<0 \big) \, .
\end{eqnarray}
In Fig.~\ref{fig:phi-dot}, the dependency of the scalar field terminal velocity to the initial electric field energy density (controlled by the parameter $\kappa$) is shown.

%%%%%%%%%%%%%%%%%%%%%%%%%%%%%%%%%%%%%%%%%%%%%%%%%%%%%%%%%%%%%%%%%%%%%%%%%%%%%%%%%%%%%%%%%%%%%%%%%%%%%%%%%%%%%%%%%%%%%%%%%%%%%%%%%%%%%%%%%%%%%%%%%%
%%%%%%%%%%%%%%%%%%%%%%%%%%%%%%%%%%%%%%%%%%%%%%%%%%%%%%%%%%%%%%%%%%%%%%%%%%%%%%%%%%%%%%%%%%%%%%%%%%%%%%%%%%%%%%%%%%%%%%%%%%%%%%%%%%%%%%%%%%%%%%%%%%

Below we estimate the mean value of the scalar field. Assuming  $\epsilon_H \approx {\rm const.}$ (which is a consistent approximation in the slow-roll limit)  we obtain
\begin{eqnarray}
\big \langle \chi\left(N\right)\big \rangle = \chi_0 -\sqrt{2\epsilon_V}\left(N-I\int_{0}^{N}\langle{\cal E}^2\left(s\right) \rangle ~{\rm d}s \right) \,.
\label{chi_aver}
\end{eqnarray}
Here the integrand has different behaviour according to the initial values of the electric field:

\begin{itemize}

\item \textbf{$\kappa \leq 0$:}\\
In this case with  ${\cal E}_{\rm ini} \leq \sqrt 2$ and with  $\langle{\cal E}^2 \rangle$ given in Eq.~\eqref{E2-averb+}, the integral of  Eq. \eqref{chi_aver} yields, 
\begin{eqnarray}
\big \langle \chi\left(N\right)\big \rangle = \chi_0 - \sqrt{2\epsilon_V}N \left( 1-I\mathcal{E}_{\rm ini}^2-\dfrac{I}{2}\left( 2b\mathcal{E}_{\rm ini}^2+3D^2 \right)N \right) \, .
\end{eqnarray}

\item \textbf{$\kappa \rightarrow -\infty$:}\\
This corresponds to the case  in which the initial electric field amplitude is equal to the  classical attractor initial condition (${\cal E}_{\rm ini} = 1$) and the system evolves as a Wiener process with no drift.
Substituting  Eq.~\eqref{E-averb0} into Eq. \eqref{chi_aver} we obtain
\begin{eqnarray}
\label{chival-1}
\big \langle \chi\left(N\right)\big \rangle = \chi_0 - \sqrt{2\epsilon_V}N \left( 1-\dfrac{3ID^2}{2}N \right) \,.
\end{eqnarray} 
We see that  the dependency of $ \langle \chi  \rangle$ to the initial value of the electric field  has disappeared when the system evolves as a Wiener process with no drift.

\item \textbf{$\kappa>0$:}\\
    Assuming that ${\cal E}_{\rm ini} \ge \sqrt 2$ and the system is in its equilibrium state of electric field distribution, the expectation  value of the scalar field is given by
\begin{eqnarray}
\label{chival-2}
\big \langle \chi\left(N\right)\big \rangle = \chi_0 - \sqrt{2\epsilon_V} N \left(1-\dfrac{3ID^2}{2|b|} \right) \, ,
\end{eqnarray}
where Eq.~\eqref{E-averbmin} has been  used.
\end{itemize}

\section{Summary and Discussions}
\label{Summary}

In this work we have revisited the stochastic effects in the model of 
$f^2F^2$ anisotropic inflation and derived the associated Langevin equations for the gauge field and inflaton field perturbations. 

We have found that the fate of the electric field energy density in the presence of stochastic noises depends to some extend  on the initial conditions.  We have introduced the parameters $q$ and $\kappa$  in  which the former is a measure of the level of anisotropy (related to the initial parameter $c$)  while the latter is a measure of the initial electric field energy density.  In the region $\kappa \leq 0$, corresponding to $0 \leq \rho_{\rm E}^{\rm ini} \leq  \sqrt 2 \rho_{\rm E}^{\rm att}$, the electric field components evolve more or less like a Brownian motion. Specifically,   the variance of the electric field energy density  increases linearly with the number of e-fold $N$. As a result,  the  near isotropy condition $R_E \ll 1$ is spoiled only after a large number of e-fold has elapsed,   $N \simeq 10^q - 10^{10}$. %( $ N\sim 10^{q}$ with $q \sim 7$).  
The situation is more interesting when $\kappa >0$, corresponding to $\rho_{\rm E}^{\rm ini} > \sqrt 2 \rho_{\rm E}^{\rm att}$.  Despite the fact that for $0 < \kappa \lesssim 1.5$ the stochastic effects can not be ignored in one number of e-fold,  but in all  regions of $\kappa>0$ the stochastic force is balanced by the classical force. Consequently,  the probability density of electric field reaches a stationary state and the velocity of the scalar field  approaches a terminal velocity. Hence, we suggest that the classical attractor mechanism of  \cite{Watanabe:2009ct}  with the electric field Eq. \eqref{E_att} is replaced by its stationary value Eq. \eqref{E-averbmin}. The mean-reversion phenomenon of Ornstein-Uhlenbeck process drives the  system to  the equilibrium state. This is similar to the classical feedback mechanism driving the system to its classical attractor regime, but considerably faster. The violation of isotropy  condition $R_E \sim 1$ yields $0<2 \kappa \lesssim{q-11}$ which can not be met for realistic range of parameter $q$, say $q \sim 7$.  As a result, when $\kappa>0$, the predictions of the $f^2F^2$ inflationary model can be consistent with cosmological observations  if  the system is in  equilibrium regime by the time $N=N_{\rm CMB}$ and the background electric field energy is given by its equilibrium value.  This is one important difference of our analysis compared to the results of \cite{ Fujita:2017lfu}.

Another  non-trivial result,  as shown in Fig. \ref{fig:E-N}, is that  for $E_{\rm ini} > E_{\rm att}$ the system falls  into its stationary state much faster than the time required  for  the classical system to reach to its attractor regime.  In the spacial case  $E_{\rm ini} = E_{\rm att}$, the stochastic noises spoil the isotropy condition just when the system reaches to its classical attractor phase  \cite{Watanabe:2009ct}. 
Finally,  for $E_{\rm ini} \leq E_{\rm att}$, the electric field energy density grows continuously and the system does not reach to an equilibrium stage. As shown in Fig. \ref{fig:RE-N}, the number of e-folds when the near isotropy condition is violated is larger than the corresponding  number of e-folds in classical system in the absence of noises.

Our findings partially disagree with those of  \cite{Fujita:2017lfu} who have found  
that  the probability for the  statistical anisotropy to be consistent with the observational bounds  is around $10^{-3}\%$. In addition, they concluded that  this result is independent of the parameter $c$ and the initial value of the electric field. In the contrary, we have shown here that this probability is not small if the initial value of the electric field is larger than the classical attractor one $(E_{\rm ini} > E_{\rm att})$.  The details of this  probability depends on the space of the parameters  $(q,\kappa)$ and can approach to $100\%$.

Finally,  we also  have obtained the  Langevin equation associated with the scalar field 
and have calculated the expectation values of  the field and its velocity. We have shown that in the equilibrium state the velocity of the scalar field approaches to a terminal velocity which depends on the initial value of the electric field. The existence of a constant terminal velocity for the scalar field in the slow-roll regime reflects the fact that the scalar field falls into its attractor regime when the electric field is in its equilibrium state.

\vspace{1cm}

 %%%%%%%%%%%%%%%%%%%%%%%%%%%%%%%%%%%%%%%%%%%%%%%%%%%%%
 
 {\bf Acknowledgments:}  A. T. would like to thank  Saramadan (Iran Science Elites Federation) for support.

%%%%%%%%%%%%%%%%%%%%%%%%%%%%%%%%%%%%%%%%%%%%%%%%%%%%%
\appendix
\section{Noise Terms Correlation Functions}
\label{Noise}

The correlation of electric field noises is given by
\begin{eqnarray}
\langle \sigma^E_i (t,\mathbf{x}) \sigma^E_j (t',\mathbf{x'}) \rangle &=& \varepsilon^2 H^4 \int \frac{d^3k} {\left(2\pi\right)^3} \int \frac{d^3k'} {\left(2\pi\right)^3} \,a(t)a(t') e^{i\mathbf{k}.\mathbf{x}}e^{i\mathbf{k'}.\mathbf{x'}} \delta \left(k-\varepsilon aH\right) \delta \left(k'-\varepsilon a'H\right)
\nonumber\\&&~~~~~~~~~~~~~~~~~~~~~~~~~~~~\times
\langle E_i(t,{\mathbf{k}}) E_j(t',{\mathbf{k'}}) \rangle
\nonumber\\&=&
\varepsilon^2 H^4   \frac{a(t)a(t)} {\left(2\pi\right)^{3}} (\varepsilon aH)^2 \dfrac{\delta \left(t-t'\right)}{\varepsilon aH^2}  E_\lambda(t,k_c)\,E_{\lambda}^{*}(t,k_c)~
\nonumber\\&&~~~~~~~~~~~~~~~~~~\times
\int e^{ik_c|\mathbf{x-x'}|\cos\theta} (\delta_{ij}-\hat{k}_i \hat{k}_j) ~\mathrm{d}\Omega \,,
\end{eqnarray}
where $k_c \equiv \varepsilon a H$ and we have used Eq. \eqref{ee}. At a fixed spatial point $\mathbf{x} \rightarrow \mathbf{x}'$, the correlation is simplified to 
\begin{eqnarray}
\langle \sigma^E_i (t,\mathbf{x}) \sigma^E_j (t',\mathbf{x}) \rangle &=& 
\varepsilon^3 H^4 a^3 \dfrac{|E_\lambda(k_c)|^2}{3\pi^2} \delta_{ij} \delta(t-t')\,,
\end{eqnarray}
 
Using Eq. \eqref{E_superhorizon}, the correlation function at large-scale, $-k_c \eta = \varepsilon \rightarrow 0$, is given by
\begin{eqnarray}
\langle \sigma^E_i \sigma^E_j \rangle &=& 
{3 H^5 \over 2 \pi^2} \delta_{ij} \delta(t-t') \,,
\end{eqnarray}
which can be rewritten as Eq. \eqref{sigma-sigma}. Also the correlation of the momentum noises at a fixed spatial point can be calculated in the same way, 
\begin{eqnarray}
\langle \tau^E_i (t,\mathbf{x}) \tau^E_j (t',\mathbf{x}) \rangle &=&
\varepsilon^3 H^4 a^3 \dfrac{|E'_\lambda(k_c)|^2}{3a^2\pi^2} \delta_{ij} \delta(t-t')\,,
\label{tauE}
\end{eqnarray}
where $E'_\lambda = \partial_\eta E_\lambda$. Hence on super-horizon scales, the correlation of the momentum noise  is given by
\begin{eqnarray}
\langle \tau^E_i (t) \tau^E_j (t') \rangle &=&
- \varepsilon^4   \dfrac{H^7}{6\pi^2} \delta_{ij} \delta(t-t') \,,
\label{tauE2}
\end{eqnarray}
where \eqref{E_superhorizon} is used. From this result we see that on large scales $\tau \sim {\cal O}(\varepsilon^2)$.

%%%%%%%%%%%%%%%%%%%%%%%%%%%%%%%%%%%%%%%%%%%%%%%%%%%%%
\section{Probability Distribution Function}
\label{PDF}
Suppose we are given a random variable $X$ with probability distribution function (PDF) $f_X(x)$ and accumulative distribution function $F_X(x)$ which is defined as follows
\begin{equation}
    F_X(X=x)=P(X<x)=\int_{x'<x}f_X(x')dx' \,.
\end{equation}
 Now let $f_Y$ be the PDF and $F_Y$ be the accumulative distribution function of $Y=g(X)$, where $g(x)$ is a real function. We then have
\begin{equation}\label{FY}
    F_Y(Y=y)=P(Y<y)=P(g(X)<y)=\int_{y'<y}f_Y(y')dy' .
\end{equation}
Calculating the PDF of $g(X)$ needs the knowledge about the general behavior of $g(X)$ and determining the domain of $x$ in which $g(x)<y$. We restrict ourselves to the case used in the paper. In other words we set $g(X)=X^2$. Hence we have
\begin{equation}\label{Fyx^2}
      F_Y(Y=y)=P(-\sqrt{y}<x<\sqrt{y})=\int^{\sqrt{y}}_{-\sqrt{y}}f_X(x)dx.
\end{equation}
From the   definition of the accumulative distribution function  one can easily see that $f_Y(y)=\frac{dF_Y(y)}{dy}$. Therefore for \eqref{Fyx^2} we have
\begin{equation}\label{finalpdf}
    f_Y(y)=\frac{f_X(\sqrt{y})}{\sqrt{y}} \, .
\end{equation}
 This is the expression we have used to  determine the PDF of ${\cal E}_i^2$ from  ${\cal E}_i$, as discussed in the main draft. 
 
 The next important expression is the PDF of the sum of two random variables. Suppose $X$ and $Y$ are two random variables. We would like to determine the PDF of $Z=X+Y$. Suppose $F_Z(z)$ be the accumulative distribution and $f_Z(z)$ be the PDF of $Z$. Then we have
 \begin{equation}
    F_Z(z)=P(Z<z)=P(X+Y<z)=\int_{z'<z}f_Z(z')dz' \, .
 \end{equation}
 On the other hand,  we can write
 \begin{equation}\label{cond1}
    P(X<z-Y)=\int_YP(X<z-y|Y=y)f_Y(y)dy,
 \end{equation}
 where $P(X<z-y|Y=y)$ is the conditional probability of $X+Y<z$ if $Y=y$. Furthermore,  one can write
 \begin{equation}\label{cond2}
     P(X<z-y|Y=y)=\int_{x'<z-y}f_X(x')dx' \, .
 \end{equation}
 So from Eqs. \eqref{cond1} and \eqref{cond2} one can deduce that
 \begin{equation}\label{accumsum}
     F_Z(z)=\int_Y\left(\int_{x'<z-y}f_X(x)dx'\right)f_Y(y)dy \, .
 \end{equation}
 By taking the derivative of \eqref{accumsum} we get the following equation for PDF of $z$ which is the convolution of $f_X$ and $f_Y$:
 \begin{equation}
     f_Z(z)=\int_Yf_X(z-y)f_Y(y)dy \, .
 \end{equation}
 Now suppose we are given two independent random variables $Y_1$ and $Y_2$ whose PDF is given by Eq. \eqref{finalpdf}. Note that $Y_1$ and $Y_2$ are two positive variables and so $P(Y_2<z-y_1|L=y_1>z)=0$. Therefore we can write Eq. \eqref{accumsum} for $Z=Y_1+Y_2$ as
 \begin{equation}
     f_Z(z)=\int_0^zf_{Y_2}(z-y_1)f_{Y_1}(y_1)dy_1= \int_0^z\frac{f_{X_2}(\sqrt{z-y_1})f_{X_1}(\sqrt{y_1})}{{\sqrt{y_1(z-y_1)}}} \, dy_1 \, .
 \end{equation}
 
 By repeating the above process one can obtain the PDF for sum of any arbitrary number of random variables. For three random variables we have
 \begin{equation}
 \begin{split}
  &f_Z(z)=\int_0^zf_{Y_1+Y_2}(z-y_3)f_{Y_3}(y_3)dy_3=\\& =\int_0^z\left(\int_0^{z-y_3}\frac{f_{X_2}(\sqrt{z-y_3-y_1})f_{X_1}(\sqrt{y_1})}{\sqrt{y_1(z-y_3-y_1)}}dy_1\right)\frac{f_{X_3}(y_3)}{\sqrt{y_3}}dy_3 
  \end{split}
 \end{equation}
 This is our starting point to derive the PDF of
 ${\cal E}^2 = \sum_{i=1}^{3} {\cal E}_i^2$.
%%%%%%%%%%%%%%%%%%%%%%%%%%%%%%%%%%%%%%%%%%%%%%%%%%%%%

\end{document}